\documentclass[10pt,twocolumn,prd,superscriptaddress,preprintnumbers,amsmath,amssymb,nofootinbib]{revtex4-2}
\usepackage{epsfig}  
\usepackage{graphicx}
\usepackage{hyperref}
\usepackage{color}
\usepackage[dvipsnames]{xcolor}
\usepackage{float}
\usepackage{amsfonts}
\usepackage{amsmath}
\usepackage{mathrsfs}
\usepackage{slashed}
\usepackage{soul}

\large

\begin{document}

\title{Majorana CP-violating phases and NSI effects in Neutrino Decay}

\author{Ashutosh Kumar Alok}
\email{akalok@iitj.ac.in}
\affiliation{Indian Institute of Technology Jodhpur, Jodhpur 342037, India}
	
\author{Neetu Raj Singh Chundawat}
\email{chundawat.1@iitj.ac.in}
\affiliation{Indian Institute of Technology Jodhpur, Jodhpur 342037, India}

\author{Arindam Mandal}
\email{mandal.3@iitj.ac.in}
\affiliation{Indian Institute of Technology Jodhpur, Jodhpur 342037, India}

\author{Bhavya Soni}
\email{soni.16@iitj.ac.in}
\affiliation{Indian Institute of Technology Jodhpur, Jodhpur 342037, India}

\date{\today} 
	
\begin{abstract}
In this work, we investigate the impact of neutrino decay in the presence of Non-Standard Interactions (NSI) along with the effects of the Majorana phase on neutrino decay in matter in the context of two-flavor neutrino oscillations. These effects are studied on neutrino oscillation probabilities $P_{\alpha \beta} \equiv  P(\nu_{\alpha} \to \nu_{\beta})$, and the difference $\Delta P_{\alpha \beta} \equiv P(\nu_{\alpha} \to \nu_{\beta}) - P(\bar {\nu}_{\alpha} \to \bar{\nu}_{\beta})$ for several accelerator and reactor neutrino experiments. We find that for $P_{\alpha \beta}$,  the influence of the Majorana phase on decay in matter can be replicated by the simultaneous presence of both decay and NSI.
However, precise measurements of the $P_{\alpha \beta}$ and $\Delta P_{\alpha\beta}$ observables have the potential to unequivocally identify the presence of the Majorana phase by discerning its effects from the concurrent presence of both decay and NSI.
\end{abstract}

\maketitle
\newpage	
\section{Introduction}

First predicted by Bruno Pontecorvo, neutrino oscillations have materialized as a plausible phenomenon, offering an explanation for the enduring Solar neutrino anomaly. The SNO and KamLAND experiments established the fact that the observed depletion of solar neutrino flux is primarily due to neutrino oscillations, disfavouring other possibilities such as decoherence and neutrino decay. Significant proof of neutrino oscillations has been recorded in a variety of experiments, encompassing a multitude of reactor and accelerator neutrino experiments as well. Numerous neutrino experiments have attained highly precise measurements of all mixing angles, as well as the mass-squared differences. Nevertheless, within the realm of neutrino oscillations, several issues remain unresolved. These encompass determining the sign of the mass squared difference, denoted as $\Delta_{31}=m_3^2-m_1^2$, and establishing the $CP$ violating phase in the leptonic sector. Additionally, the question of the potential existence of a non-zero Majorana phase remains open. 
The forthcoming long baseline neutrino oscillation experiment, DUNE, featuring an intense neutrino beam, is anticipated to facilitate the observation of $CP$ violation and offer crucial insights into the neutrino mass ordering \cite{DUNE:2015lol}. 
The JUNO experiment is also specifically designed to address the neutrino mass ordering problem \cite{JUNO:2015zny, JUNO:2021vlw}.

While the majority of existing data in the neutrino sector, derived from numerous neutrino oscillation experiments, can be explained within the framework of standard neutrino oscillation, the   possibility of Beyond Standard Model (BSM) contrinutions, such as non-standard interactions (NSI) and neutrino decay, remains viable. Indeed, these effects can function as sub-leading factors in neutrino oscillations. With current and imminent experiments in the neutrino sector reaching the precision era, there is a substantial opportunity to effectively investigate and probe these effects.

The domain of physics beyond the SM, often referred to as new physics, possesses the capability to modify the interactions of neutrinos with matter, consequently influencing the intricate patterns of neutrino oscillations. These alterations in neutrino oscillations induced by new physics effects, specifically termed as NSI, present an avenue for exploration in various neutrino oscillation experiments \cite{Ohlsson:2012kf,Miranda:2015dra,Proceedings:2019qno,
Farzan:2017xzy,Bakhti:2020fde,Kopp:2007ne}. The parameters governing these NSI, encapsulating the impact of physics beyond the Standard Model, manifest in the formulations of neutrino oscillation probabilities. As a result, the determination of these new physics couplings becomes achievable through a diverse array of neutrino oscillation experiments. 
Moreover, NSI can introduce supplementary sources of CP violation, a matter that has garnered renewed interest in light of the recent discrepancies observed in measurements conducted by NO$\nu$A and T2K. The NSI effects in the four Fermi interactions involving neutrinos can be incorporated within the framework of effective field theories. In the presence of NSI, the effective matter potential is modulated, and the total Hamiltonian describing the dynamics of neutrino oscillations is obtained by the addition of the vacuum, matter and NSI Hamiltonian. 

Another sub-dominant effect is the possibility of neutrino decay \cite{Bahcall:1972my,Acker:1991ej,Acker:1993sz}. Heavier eigenstates of neutrinos can decay into lighter eigenstates, accompanied by some BSM particles. The essence of neutrino decay can be captured in terms of the mass and lifetime of the neutrino mass eigenstate as $e^{-\frac{m}{\tau}\frac{L}{E}}$.  The lifetime of neutrinos is constrained by different experimental observations. Lower bound on the lifetime of $\nu_1$ is obtained to be $\tau_1/m_1 > 4 \times 10^{-3}$  s/eV \cite{Berryman:2014qha}. The lower limits on $\tau_2/m_2$ extracted from solar neutrino data and reactor neutrino experimental data are $\tau_2/m_2 > 8.7\times 10^{-4}$ s/eV and $\tau_2/m_2 > 8.7\times 10^{-4}$ s/eV, respectively \cite{Bandyopadhyay:2002qg,Berryman:2014qha}. A bound on $\nu_3$ lifetime, $\tau_3/m_3 > 2.9\times 10^{-10}$ s/eV is achieved from long baseline and atmospheric neutrino experimental data \cite{Gonzalez-Garcia:2008mgl}. Considering the decay of at least one neutrino eigenstate $\tau/m > 10^{5}$ s/eV is obtained from the supernova neutrinos \cite{Frieman:1987as}. The decay of neutrinos is also studied in the context of T2K \cite{Gago:2017zzy}, MINOS \cite{Gago:2017zzy}, and DUNE \cite{Ascencio-Sosa:2018lbk,deGouvea:2015ndi} experiments. The treatment of neutrino oscillation, taking into consideration the possibility of neutrino decay, is somewhat non-trivial. This is due to the fact that the decay Hamiltonian is not always aligned with free Hamiltonian in the mass basis \cite{Chattopadhyay:2021eba,Berryman:2014yoa,Chattopadhyay:2022ftv}.
A novel technique is instigated in \cite{Chattopadhyay:2021eba} to obtain an analytical expression for the neutrino oscillation probabilities while they are assumed to decay.    In this case, the Majorana phase becomes relevant and can appear in the probability expressions, which is not the case in normal oscillation probabilities.
 The impact of the Majorana phase on neutrino decay in a vacuum was explored in the context of two-flavour neutrino oscillation in \cite{Dixit:2022izn}  and it was shown that the effect of the Majorana phase could be captured in this framework as well. This stands in stark contrast to typical 2-flavor neutrino oscillations, where the Majorana phase, despite its existence, does not influence the oscillation probabilities. However, in scenarios involving neutrino decay, the relevance of the Majorana phase in 2-flavor oscillations emerges, opening up possibilities to investigate its existence and making the study of its effects in these sub-leading scenarios particularly fascinating. Motivated by this fact, this analysis specifically focuses on 2-flavor neutrino oscillations \footnote{The analysis in this work can be extended to three-flavor neutrino mixing, where the sub-dominant contributions such as NSI and decay  to standard neutrino oscillation can be further modified. However, this will necessitate a more rigorous examination and can be considered for future analysis.}.

Both NSI and neutrino decay are sub-leading effects that can influence neutrino oscillations at the same new physics scale. Typically, these effects are considered individually. However, in this work, we explore the possibility that both can coexist by investigating neutrino decay in the presence of NSI effects. 
We also conduct the first analytical investigation of the influence of the Majorana phase on neutrino decay in matter. 
These effects are examined in the context of neutrino transition probabilities $P(\nu_{\alpha} \to \nu_{\beta})$ and the corresponding difference observables $ P(\nu_{\alpha} \to \nu_{\beta}) - P(\bar{\nu}_{\alpha} \to \bar{\nu}_{\beta})$. The investigation is carried out across various experimental setups involving both accelerator and reactor neutrinos.

The structure of the work is as follows: In the upcoming section, we present theoretical expressions for neutrino oscillation probabilities in the context of neutrino decay in the presence of NSI. Subsequently, in Section \ref{decay-majorana}, theoretical expressions are derived for neutrino decay in matter, considering the existence of a non-zero Majorana phase. Section \ref{res} is dedicated to the analysis of these effects across various experimental setups related to neutrino oscillations. We finally conclude in Sec \ref{concl}.


\section{Neutrino decay in the presence of NSI}
\label{nsi-decay}

The oscillation of neutrinos, driven by the interplay of neutrino mixing and non-zero masses, results in the transition between different flavours over time. In the vacuum, this dynamic evolution is governed by the following Hamiltonian,
\begin{equation}
    H = \frac{1}{2E}\begin{pmatrix}
        m_{1}^{2} & 0 \\
        0 & m_{2}^{2}
    \end{pmatrix}.
\end{equation}
The mixing of the mass eigenstates is represented by the following mixing matrix,
\begin{equation}
    U = \begin{pmatrix}
      \cos{\theta} & \sin{\theta}\\
      -\sin{\theta} & \cos{\theta}
    \end{pmatrix},
\end{equation}
where $\theta$ being the mixing angle.
The oscillation probability in a vacuum is then given by,
\begin{equation}
    P_{\alpha\beta} = \sin^{2}{(2\theta)}\sin^{2}{\left(\frac{\Delta m^{2}L}{2E}\right)},
    \label{prob-vac}
\end{equation}
where $\Delta m^{2} = m_{2}^{2} - m_{1}^{2}$. 
It is evident that the oscillation probability depends on two intrinsic properties of neutrinos: the mixing angle $\theta$ and the mass-squared difference of the neutrino mass eigenstates.

The interaction of neutrinos with matter can alter the effective values of these parameters. The modified values of mass-squared difference and mixing angle in matter are given as follows
\begin{equation}
    \Delta m_{m}^{2} = \sqrt{(\Delta m^{2} \cos{2\theta} - A)^{2} + (\Delta m^{2}\sin{2\theta})^2},
\end{equation}
and 
\begin{equation}
    \theta_{m} = \frac{1}{2}\arcsin\left[{\frac{\Delta m^{2}\sin{(2\theta)}}{\Delta m_{m}^{2}}}\right],
\end{equation}
where $A = 2\,E\,V_{CC}$ corresponds to the potential due to the charged current interaction of the neutrinos.

In addition to the matter potential under consideration here, neutrinos can also encounter a potential generated by NSI. The Lagrangian corresponding to NSI is commonly written as,
\begin{equation}
    \mathcal{L}_{NSI} = -2\sqrt{2}G_{F}\varepsilon^{ff'}_{\alpha\beta}(\overline{\nu}_{\alpha}\gamma^{\mu}P_{L}\nu_{\beta})(\overline{f}\gamma_{\mu}P_{h}f'),
\end{equation}
where $\varepsilon$'s are known as the NSI parameters, which encapsulate the effects of new physics. Usually, matter consists of neutrons, protons and electrons. Thus $f,f' = e,u,d$. The handedness is denoted by $h = L,R$. This NSI adds an additional component to the Hamiltonian, which further modifies the value of the mass-squared difference and the mixing angle. These parameters in the presence of NSI are given by \cite{Ohlsson:2012kf},
\begin{eqnarray}
   ( \Delta\tilde{m}^2 )^2&=& [\Delta m^{2} \cos{(2\theta)} - A(1 + \varepsilon_{\alpha\alpha} 
   - \varepsilon_{\beta\beta})]^{2} 
   \nonumber\\
   &&+ [\Delta m^{2} \sin{(2\theta)} + 2A\varepsilon_{\alpha\beta}]^{2},
    \label{mass-nsi}
\end{eqnarray}
and
\begin{equation}
    \tilde{\theta} = \frac{1}{2}\arcsin{\left[\frac{\Delta m^{2} \sin{(2\theta)} + 2A\varepsilon_{\alpha\beta}}{\Delta\tilde{m}^2}\right]}.
    \label{angle-nsi}
\end{equation}
In this case, the oscillation probability is given by Eq. \ref{prob-vac} with these modified values of the parameters.

The non-zero masses of neutrinos not only lead to neutrino oscillations but can also introduce the possibility of decay from heavier eigenstates to lighter ones. In the presence of neutrino decay, the Hamiltonian governing the evolution of neutrinos is given by \cite{Chattopadhyay:2021eba},
\begin{equation}
    \mathcal{H}_m = H_m - i\Gamma/2,
\end{equation}
where $H_m$ and $\Gamma$ are the free Hamiltonian and decay matrix in the mass basis, respectively. 
In general, $[H_m, \Gamma]\neq 0$, which means that we cannot express $e^{-i \mathcal{H}_m t}$ as $e^{-\Gamma t/2}e^{-iH_{m}t}$. Consequently, to derive the evolution of neutrinos, $\nu(t) = e^{-i\mathcal{H}_{m} t}\nu(0)$, we \textbf{cannot} simply employ $\nu(t) = e^{-\Gamma t/2}e^{-iH_{m}t}\nu(0)$. Instead, the expansion of $e^{-i \mathcal{H}_m t}$ must be carried out using the Baker-Campbell-Hausdorff (BCH) formula.
In general, $\mathcal{H}_m$ can be written as \cite{Chattopadhyay:2021eba}, 
\begin{equation}
    \mathcal{H}_m = \begin{pmatrix}
    a_{1}-ib_{1} & -\frac{i\gamma}{2}e^{i\chi}\\
    -\frac{i\gamma}{2}e^{-i\chi} & a_{2}-ib_{2}
    \end{pmatrix},
\end{equation}

Here, $a_1$ and $a_2$ correspond to the eigenvalues of the Hamiltonian in matter in the absence of decay. The other parameters $b_1$, $b_2$, $\gamma$, and $\chi$ correspond to the decay matrix. Notably, the non-diagonal elements contribute to the misalignment between the decay and free Hamiltonian in the mass basis. 
 In terms of these  parameters, the survival and oscillation probabilities are given as \cite{Chattopadhyay:2021eba},
\begin{eqnarray}
    P_{\alpha\alpha} &=& \frac{e^{-(b_{1}+b_{2})t}}{2}\Bigg[(1 + | \mathscr{A}(\chi)|^{2})\cosh{(\Delta_b t)}  \nonumber\\
      && + (1 - | \mathscr{A}(\chi)|^{2})\cos{(\Delta_a t)} -2 Re( \mathscr{A}(\chi)) \sinh{(\Delta_b t)}  \nonumber\\
      &&
   + 2 Im( \mathscr{A}(\chi)) \sin{(\Delta_a t)} \Bigg],
    \label{surv}
\end{eqnarray}
and
\begin{equation}
    P_{\alpha\beta} = \frac{e^{-(b_{1}+b_{2})t}}{2} |\mathscr{B}(\chi)|^{2} [\cosh{(\Delta_b t)} - \cos{(\Delta_a t)}],
    \label{oscl}
\end{equation}
respectively. Here, $\Delta_a = a_2 - a_1$, $\Delta_b = b_2 - b_1$, $\Delta_d = \Delta_a - \Delta_b$  and the mixing terms are given by,
\begin{equation}
    \mathscr{A}(\chi) = -\cos{2\theta_m} - \frac{i\gamma}{\Delta_d} \sin{2 \theta_m} \cos{\chi},
\end{equation}
and
\begin{equation}
    \mathscr{B}(\chi) = \sin{2\theta_m} - \frac{i\gamma}{\Delta_d} (\cos{2 \theta_m} \cos{\chi} + i \sin{\chi}),
\end{equation}
where $\theta_m$ is the mixing angle in the presence of matter. The mixing matrix in matter is given as,
\begin{equation}
    U_m = \begin{pmatrix}
       \cos{\theta_m} & \sin{\theta_m}\\
       -\sin{\theta_m} & \cos{\theta_m}
    \end{pmatrix}.
\end{equation}
In the presence of NSI, the mixing matrix is modified to,
\begin{equation}
     \tilde{U} = \begin{pmatrix}
       \cos{\tilde{\theta}} & \sin{\tilde{\theta}}\\
       -\sin{\tilde{\theta}} & \cos{\tilde{\theta}}
    \end{pmatrix}.
\end{equation}
Therefore, in the presence of NSI along with the potential for decay, the survival and oscillation probabilities of neutrinos can be derived from Eq. \ref{surv} and Eq. \ref{oscl}, respectively, with the adjusted values of $\Delta_a$ and $\theta$. The modified value of $\theta$ is denoted as $\tilde{\theta}$ and is determined by Eq. \ref{angle-nsi}. The value of $\Delta_a$ undergoes a modification to $\frac{\Delta\tilde{m}^2}{2E}$, obtained from Eq. \ref{mass-nsi}.

\section{Neutrino decay in matter with a Majorana phase}
\label{decay-majorana}
The nature of neutrinos, Majorana or Dirac, is an open question. To reveal the nature, many experiments have been proposed based on the detection of neutrino-less double beta decay. If neutrinos are considered to be Majorana, an extra phase $\phi$ appears in the unitary matrix in the form,

\begin{equation}
    O = \begin{pmatrix}
      1 & 0\\
      0 & e^{i\phi}
\end{pmatrix}
\end{equation}

However, in the absence of off-diagonal phases in the mass matrix, the presence of this phase does not affect the oscillation probabilities in 2-flavor neutrino oscillations because the decay eigenbasis is the same as the mass eigenbasis. Thus, this phase can only be realised when both of these basis are not the same, which is the case of neutrino decay. The off-diagonal decay matrix contains a complex phase, and in the presence of this phase and other off-diagonal parameters, the Majorana phase becomes relevant and appears in 2-flavor neutrino oscillations \cite{Dixit:2022izn}.

 The new mixing matrix containing the phase $\phi$ modifies the oscillation probabilities in the presence of decay and matter effects.  This Majorana phase and the decay phase $\chi$, collectively manifest as a modified phase: $\chi-\phi$. Thus, the new forms of $\mathscr{A}(\chi)$ and $\mathscr{B}(\chi)$ are given as,
\begin{equation}
    \mathscr{A}(\chi-\phi) = -\cos{2\theta_m} - \frac{i\gamma}{\Delta_d} \sin{2 \theta_m} \cos{(\chi-\phi)}.
\end{equation}
and
\begin{equation}
    \mathscr{B}(\chi-\phi) = \sin{2\theta_m} - \frac{i\gamma}{\Delta_d} (\cos{2 \theta_m} \cos{(\chi-\phi)} + i \sin{(\chi-\phi)}),
\end{equation}
\begin{table*}[t]
    \centering
    \begin{tabular}{|c|c|c|c|c|c|c|c|c|c|c|}
        \hline
        $\theta_{12}$ & $\theta_{13}$ & $\theta_{23}$ & $\Delta m_{21}^{2}\,\rm (eV^{2})$ & 
        $\Delta m_{23}^{2} \,\rm(eV^{2})$ &
        $\varepsilon_{ee}$ & $\varepsilon_{\mu\mu}$ & $\varepsilon_{\tau\tau}$ & $\varepsilon_{e\mu}$ & $\varepsilon_{e\tau} $& $\varepsilon_{\mu\tau}$ \\
        \hline
        \hline
        $34.3^\circ$ & $8.58^\circ$ & $48.79^\circ$ & $7.5\times 10^{-5}$ & $2.56\times 10^{-3}$&
        [0.24, 2.27] & [-0.30, 0.37] & [-0.30, 0.38] & [-0.3, 0.16] & [-0.76, 0.53] & [-0.03, 0.03]\\
        \hline
\end{tabular}
    \caption{The values of mixing angles, mass-squared differences in normal ordering and the NSI parameters \cite{Coloma:2020nhf}}
    \label{NSI}
\end{table*}
In the case of antineutrinos, the probabilities $P_{\bar{\alpha}\bar{\beta}}$ and $P_{\bar{\alpha}\bar{\alpha}}$ are derived by replacing $\phi \to -\phi$ and $\chi \to -\chi$. As $\mathscr{A}(\chi-\phi)$ exclusively involves the cosine term, this substitution has no impact on it. However, only $\mathscr{B}(\chi-\phi)$ undergoes a change due to the sine term. Consequently, only $P_{\bar{\alpha}\bar{\beta}}$ is altered. 
The presence of such a non-vanishing phase can generate $CP$-violating effects even in 2-flavor. This $CP$-violating effect can be quantified as the difference in the vacuum oscillation probabilities ($\Delta P_{\alpha\beta}  \equiv P(\nu_{\alpha} \to \nu_{\beta}) - P(\bar {\nu}_{\alpha} \to \bar{\nu}_{\beta})$) for neutrinos and anti-neutrinos, which reads,
\begin{eqnarray}
    \Delta P_{\alpha\beta} =  2 e^{-(b_{1}+b_{2})t} [\cosh{(\Delta_b t)} - \cos{(\Delta_a t)}]  && \nonumber \\   \times \sin{2\theta}\frac{\gamma\Delta_a}{|\Delta_d|^2}\sin{(\chi-\phi)}\,,
    \label{Vacuum}
\end{eqnarray} 
and $\Delta P_{\alpha\alpha} = 0$. 
\begin{figure*}[t!]
    \centering
\includegraphics[width = 0.47\textwidth]{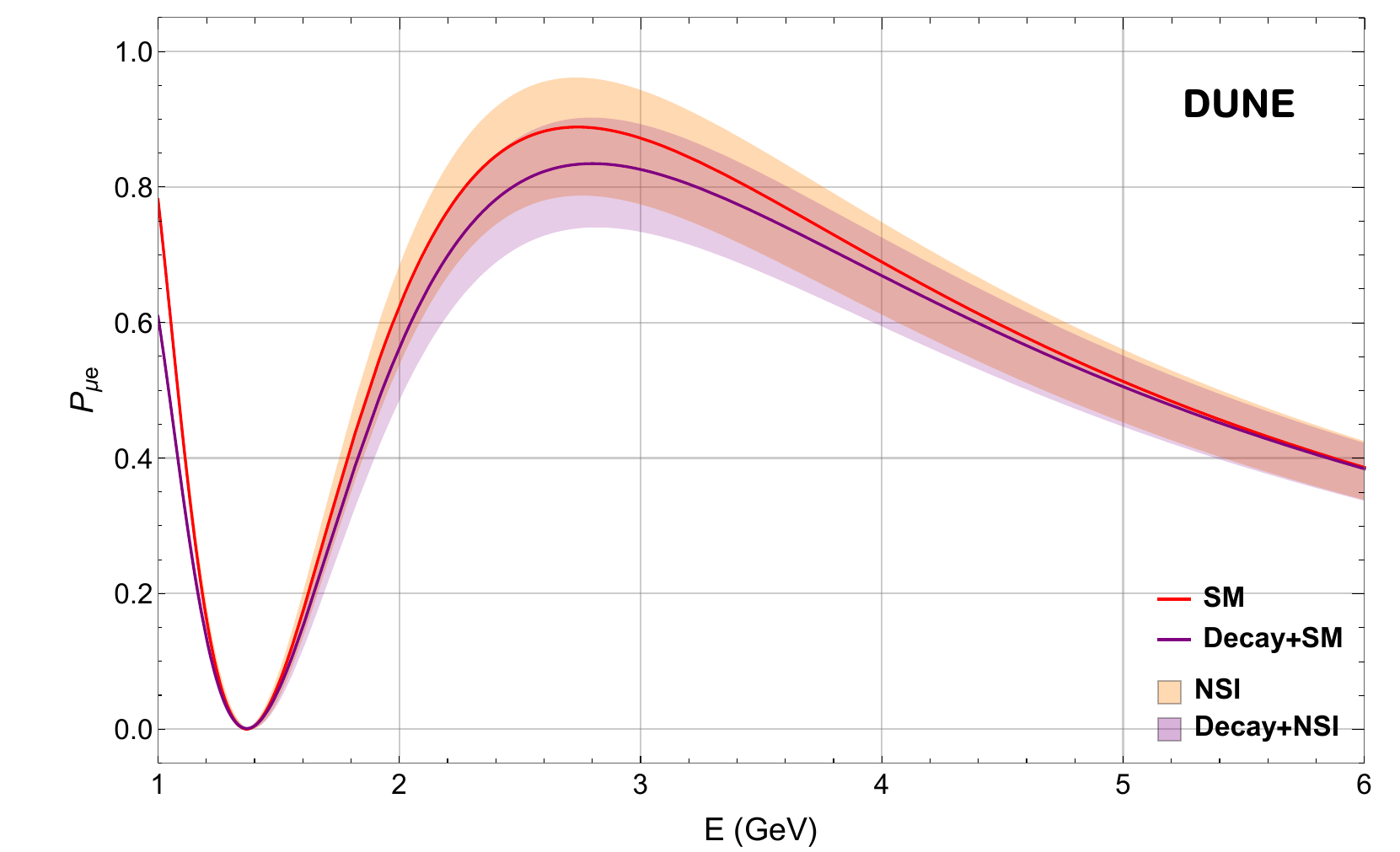}
\includegraphics[width = 0.47\textwidth]{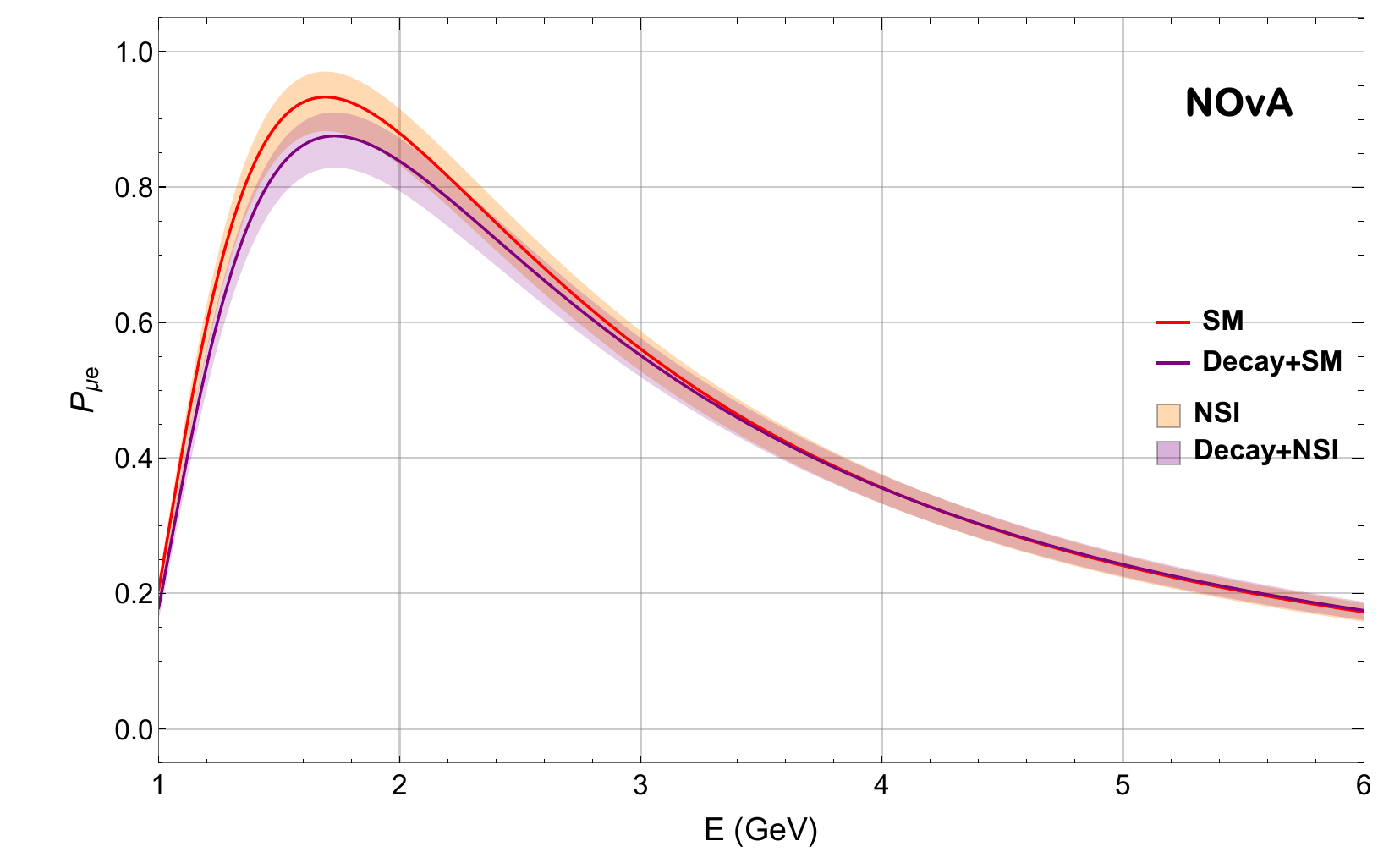}\\
\includegraphics[width = 0.47\textwidth]{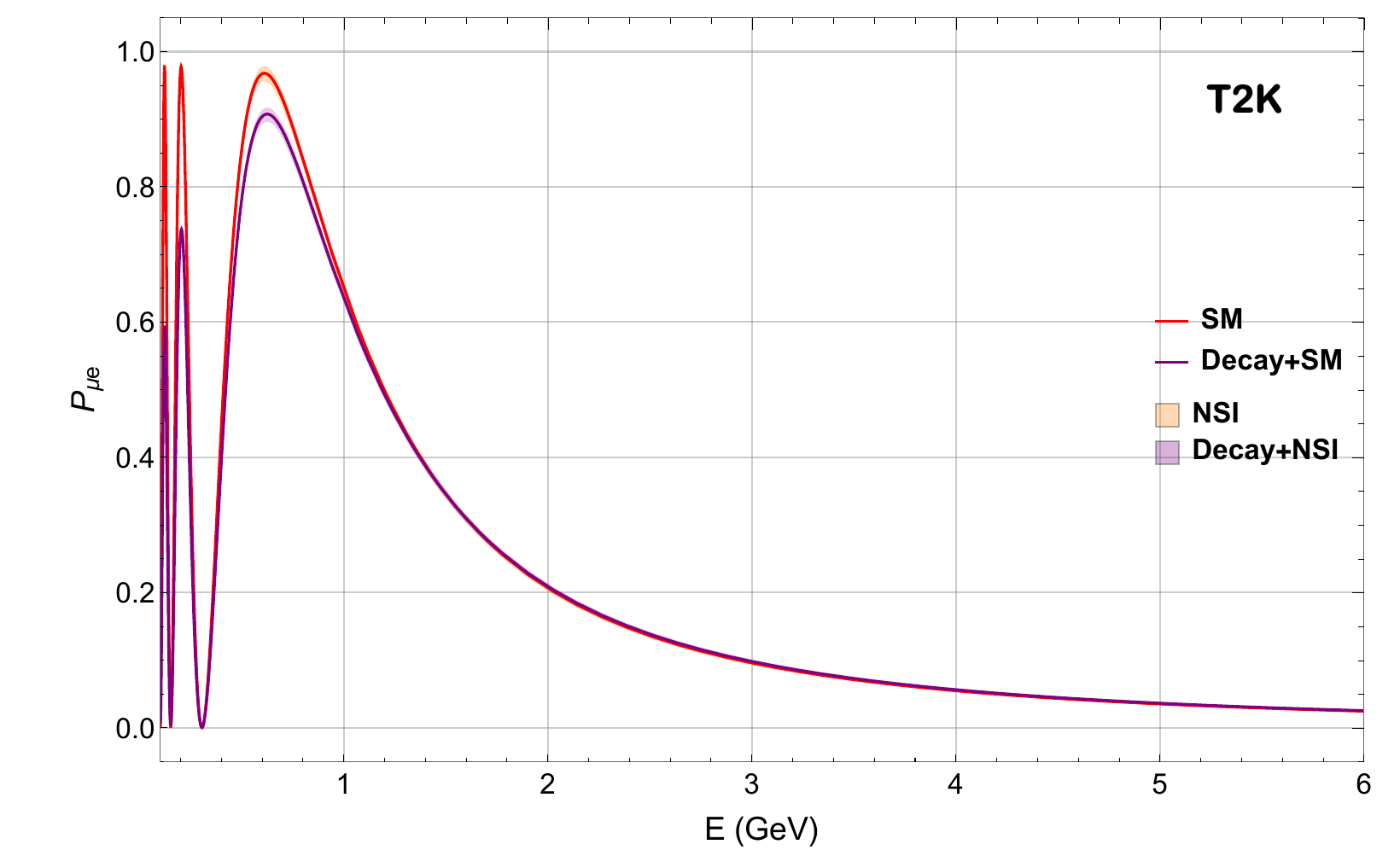}
\includegraphics[width = 0.47\textwidth]{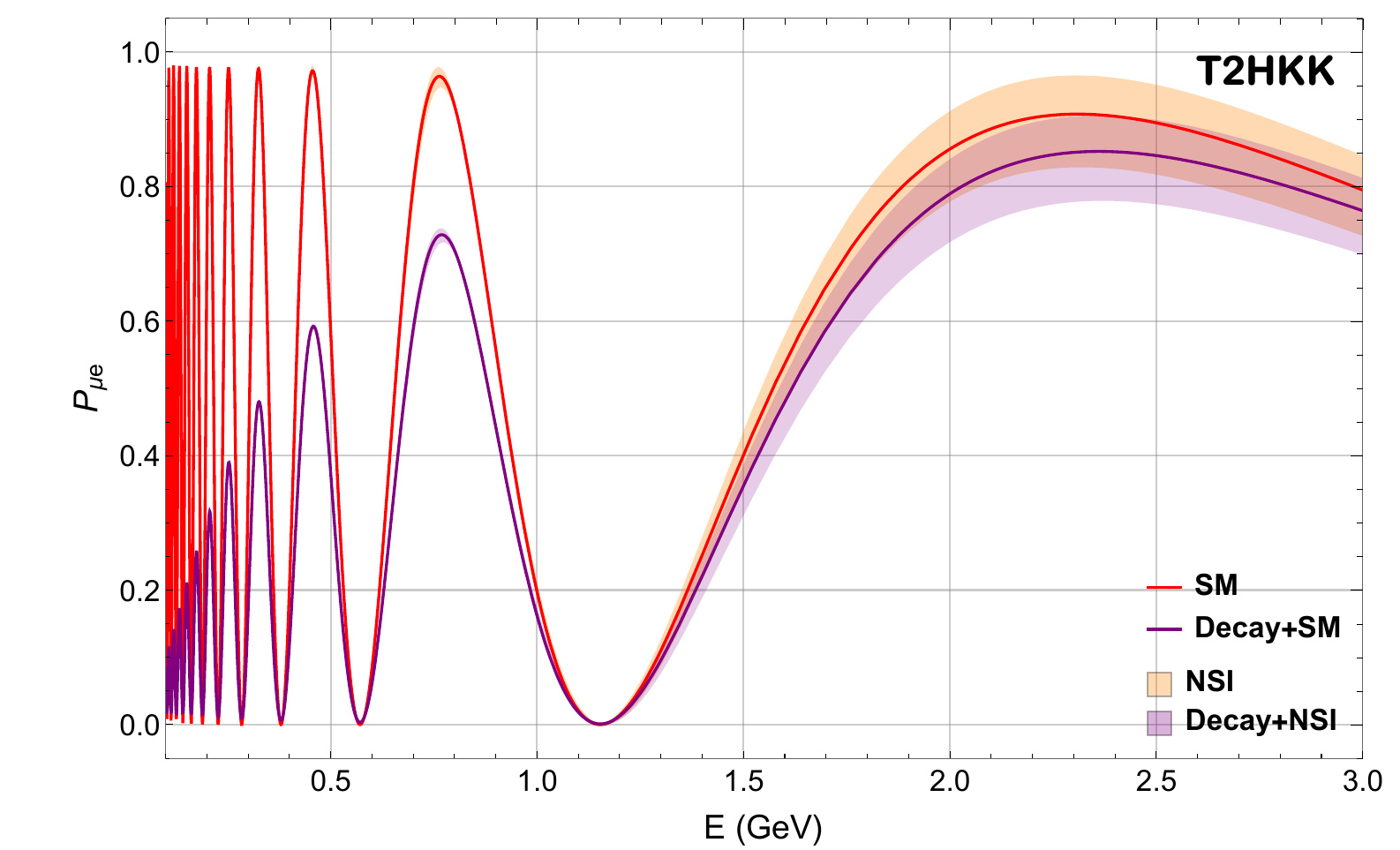}\\
\caption{Oscillation probabilities for DUNE, No$\nu$A, T2K and T2HKK accelerator experiments.
The red line illustrates neutrino oscillation in matter, and the orange band indicates potential NSI parameter effects on matter oscillation probability. The purple solid line represents decay probability in matter, with the corresponding purple band reflecting NSI impacts on neutrino decay. 
} 
\label{accelerator-wo-majorana}
\end{figure*}

\begin{figure*}[htb]
    \centering
\includegraphics[width = 0.47\textwidth]{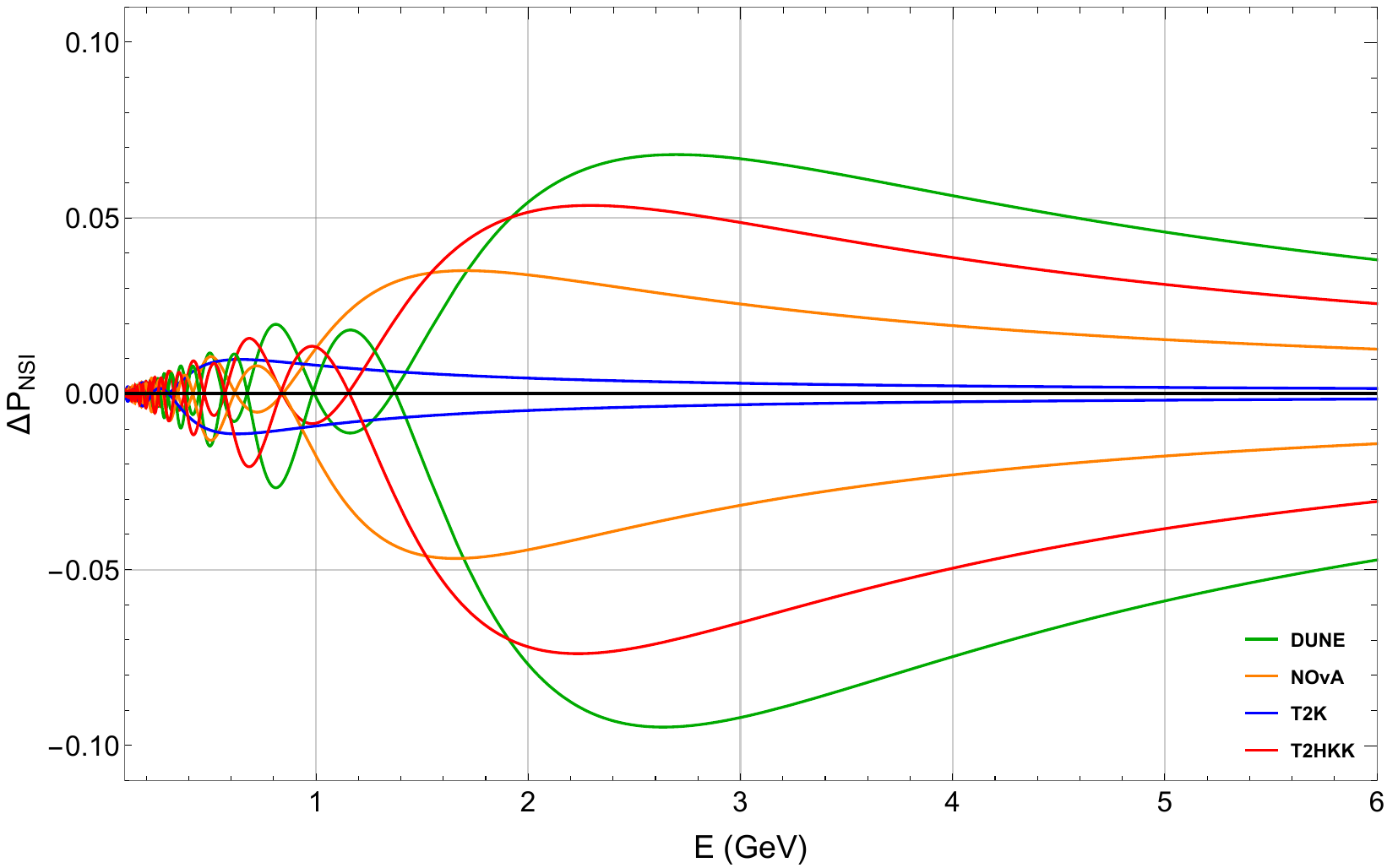}
\caption{The two solid lines illustrate the maximum deviation in $\Delta P_{\rm NSI}= P_{\mu e}^{\rm decay + NSI} - P_{\mu e}^{\rm decay + SM}$,  i.e. the maximum deviation in $P_{\mu e}$ due to the combined influences of
NSI and decay as compared to the neutrino decay in matter with SM interactions, for the  various accelerator experimental setups.  The colors green, orange, blue, and red specifically denote the DUNE, NOvA, T2K, and T2HKK experiments, respectively.} 
\label{accelerator-diffwo-majorana}
\end{figure*}

Thus, in vacuum oscillations, $CP$-violating effects vanish if (i) $\chi$ and $\phi$ are absent, (ii) the two phases are equal, or (iii) the diagonal term in the decay matrix is absent ($\gamma = 0$). In the presence of matter, on the other hand, $CP$-violating effects become more intricate due to the difference in the nature of the matter potential for neutrinos and antineutrinos. In this case, the $CP$-difference contains contributions from the potential $A$, in addition to the two phases $\chi$ and $\phi$. The resulting $\Delta P_{\alpha\beta}$ assumes the form:
\begin{eqnarray}
\Delta P_{\alpha\beta} &=&  \frac{e^{-(b_{1}+b_{2})t}}{2} \Big[\cosh{(\Delta_b t)}(|\mathscr{B}(\chi-\phi)|^{2}-|\mathscr{\Bar{B}}(\chi-\phi)|^{2}) 
\nonumber\\  && 
-\cos{(\Delta}_a t)|\mathscr{B}(\chi-\phi)|^{2}+
\cos{(\Bar{\Delta}_a t)}|\mathscr{\Bar{B}}(\chi-\phi)|^{2}\Big] \,,
 \nonumber \\
\label{MatterBhavya}
\end{eqnarray}
where $\Bar{\Delta}_a$ and $\Bar{\Delta_d}$ represent their respective differences for antineutrino and $\mathscr{\Bar{B}}(\chi-\phi)$ is given as,
\begin{equation}
    \mathscr{\Bar{B}}(\chi-\phi) = \sin{2\Bar{\theta}_m} - \frac{i\gamma}{\Bar{\Delta_d}} (\cos{2\Bar{\theta}_m} \cos{(\chi-\phi)} - i \sin{(\chi-\phi)}).
\end{equation}
The sole modification in $\mathscr{\Bar{B}}$ results from the alteration in the sign associated with the sine term, stemming from the earlier mentioned substitutions for antineutrinos. When the matter effect is considered, all terms for neutrinos and antineutrinos exhibit disparities except for $\Delta_b$. Consequently, even in the absence of the phases, a non-zero difference manifests in the oscillation probabilities. We have used this difference $\Delta P_{\alpha\beta}$ to quantify the effect of the Majorana phase on decay, distinguishing it from the impact of NSI on decay in the subsequent section.

\begin{figure*}[htb!]
    \centering
\includegraphics[width = 0.47\textwidth]{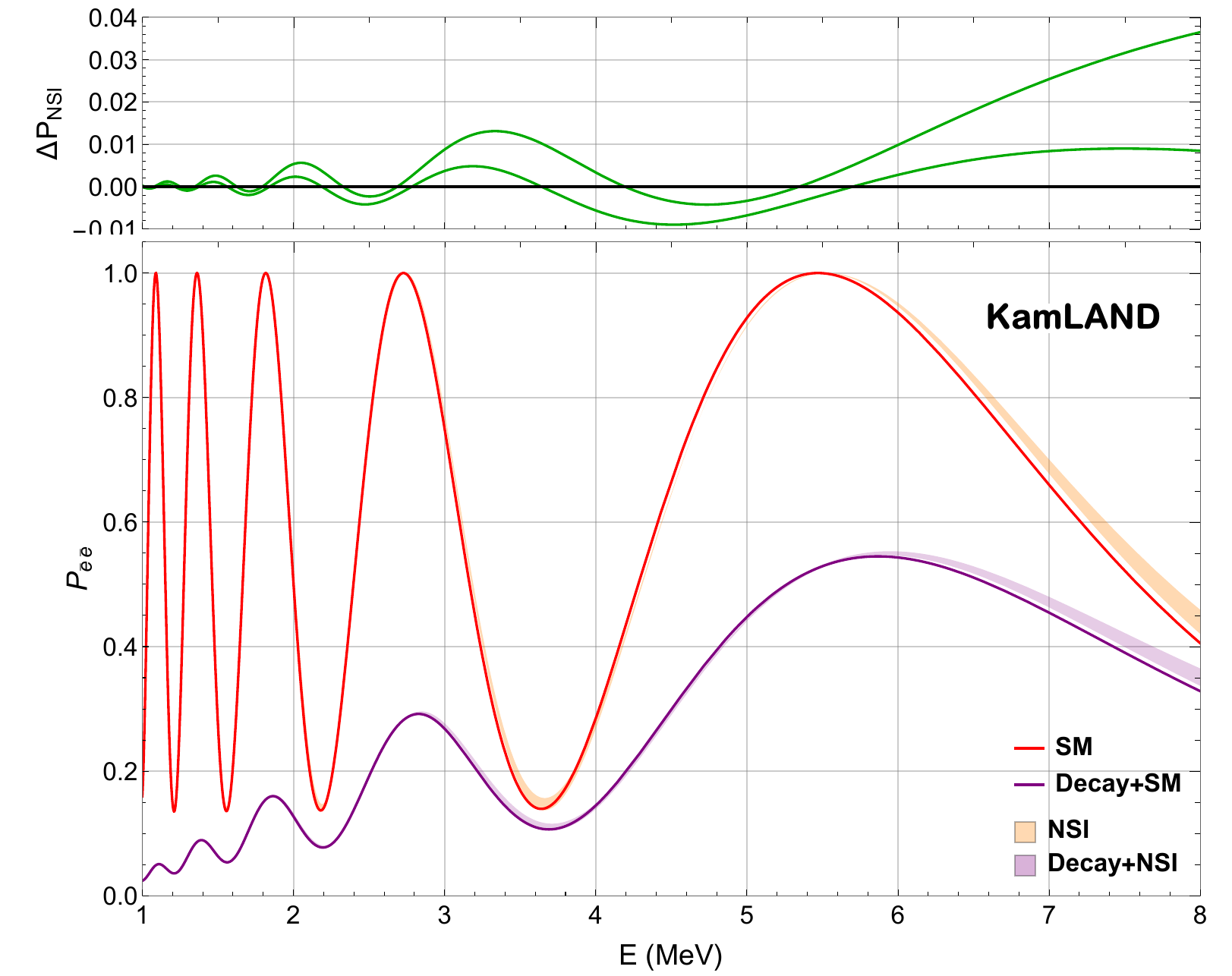}
\includegraphics[width = 0.47\textwidth]{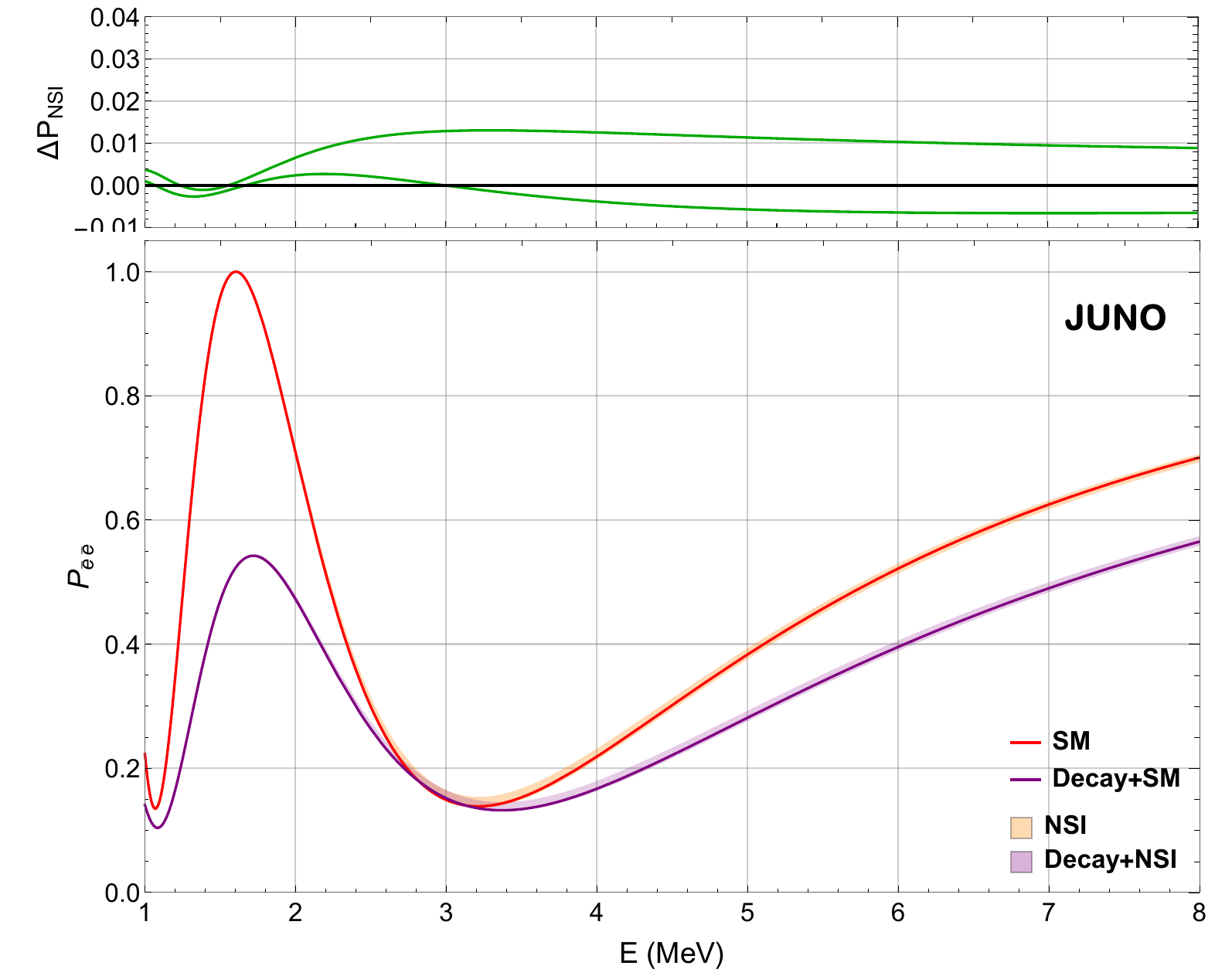}
\caption{ Survival probabilities for KamLAND and JUNO reactor experiment. Here, the plots follow the same color code and definitions as in Fig. \ref{accelerator-wo-majorana} 
}\label{reactor-wo-majorana}
\end{figure*}

\section{Results and Discussions}
\label{res}
In this section, we analyze how NSI influences the pattern of neutrino oscillations when neutrino decays in matter.  We investigate these effects across various experimental configurations associated with accelerator and reactor neutrinos.  For a particular setup, we intend to identify regions where the dual effect of decay and NSI  can be differentiated from their individual effects. We then consider the effect of the Majorana phase on neutrino decay in the matter. Finally, we consider the observable $\Delta\, P = P(\nu_{\alpha} \to \nu_{\beta}) - P(\bar{\nu_{\alpha}} \to \bar{\nu_{\beta}})$ and study the impact of dual effects of decay and NSI as well as the effect of Majorana phase on neutrino decay in matter.

\begin{figure*}[htb!] 
    \centering
    \includegraphics[width = 0.47\textwidth]{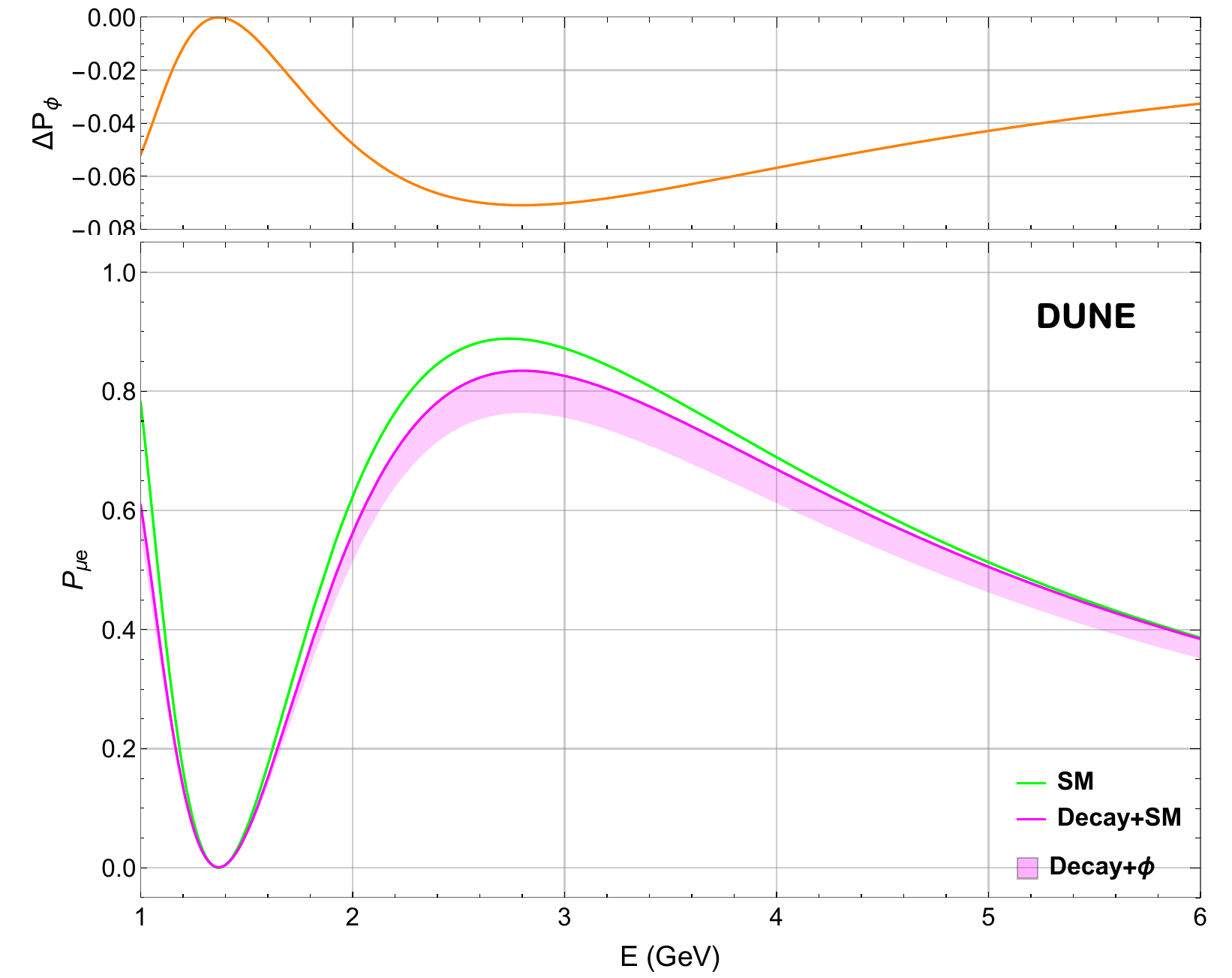}
    \includegraphics[width = 0.47\textwidth]{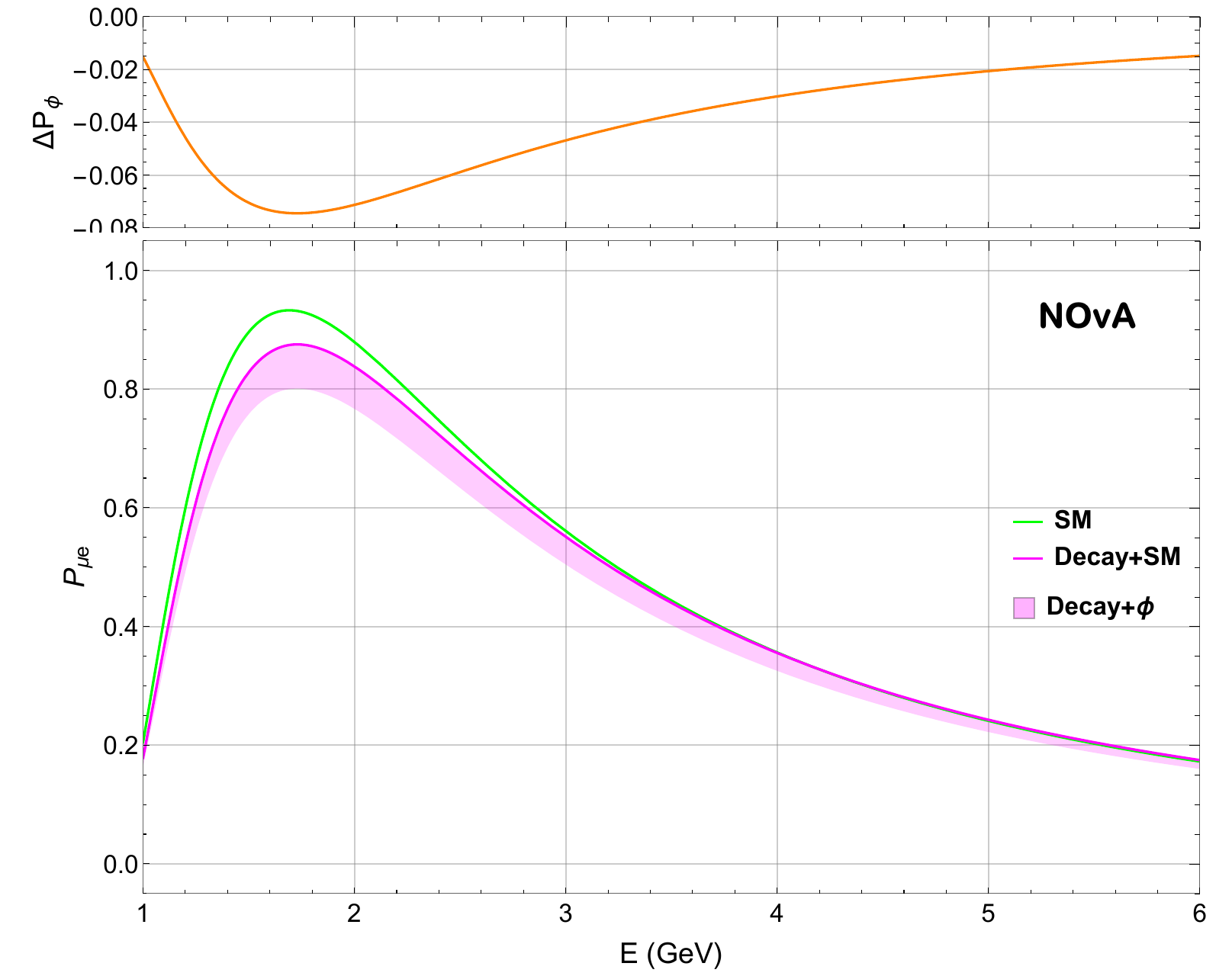}\\
    \includegraphics[width = 0.47\textwidth]{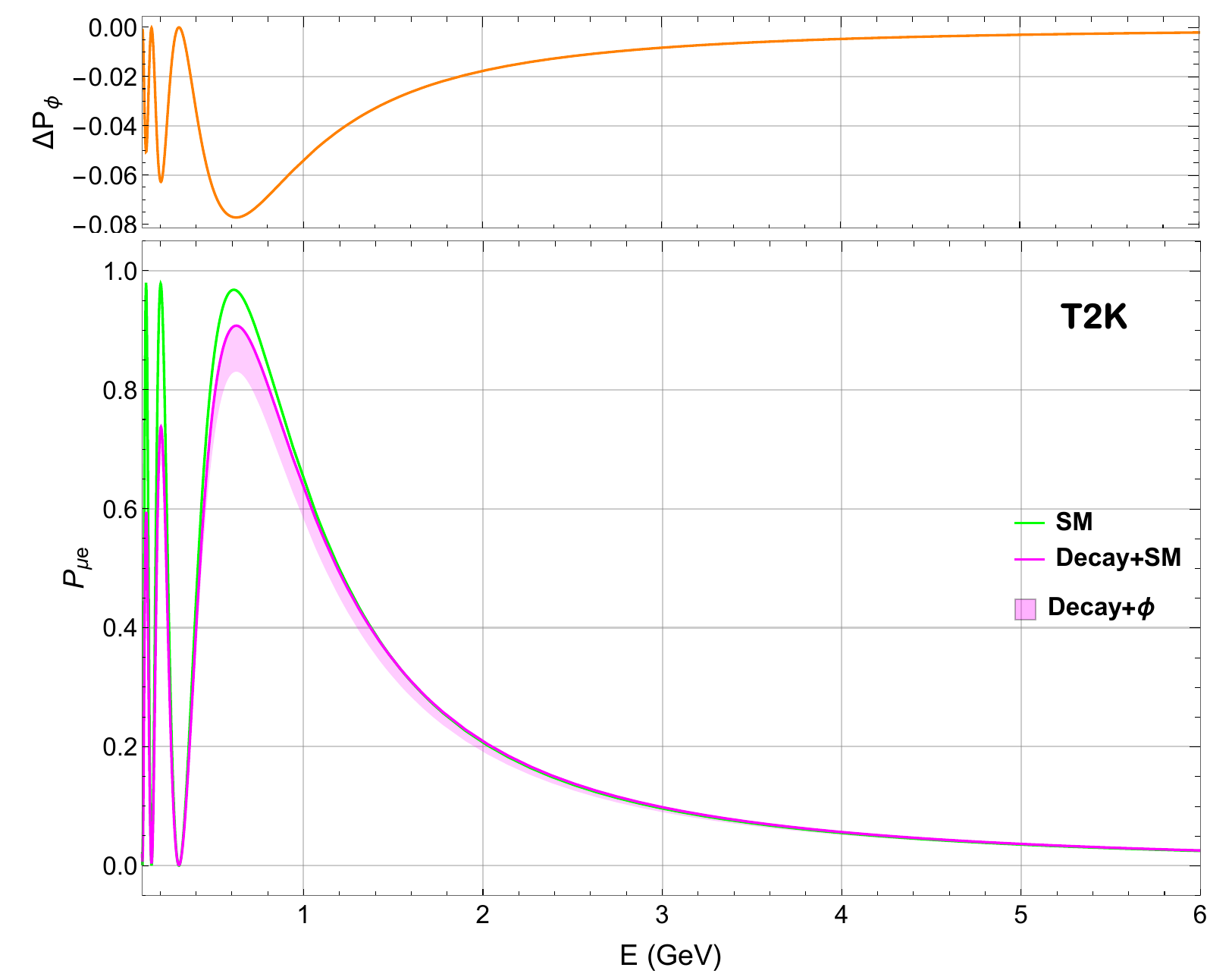}
    \includegraphics[width = 0.47\textwidth]{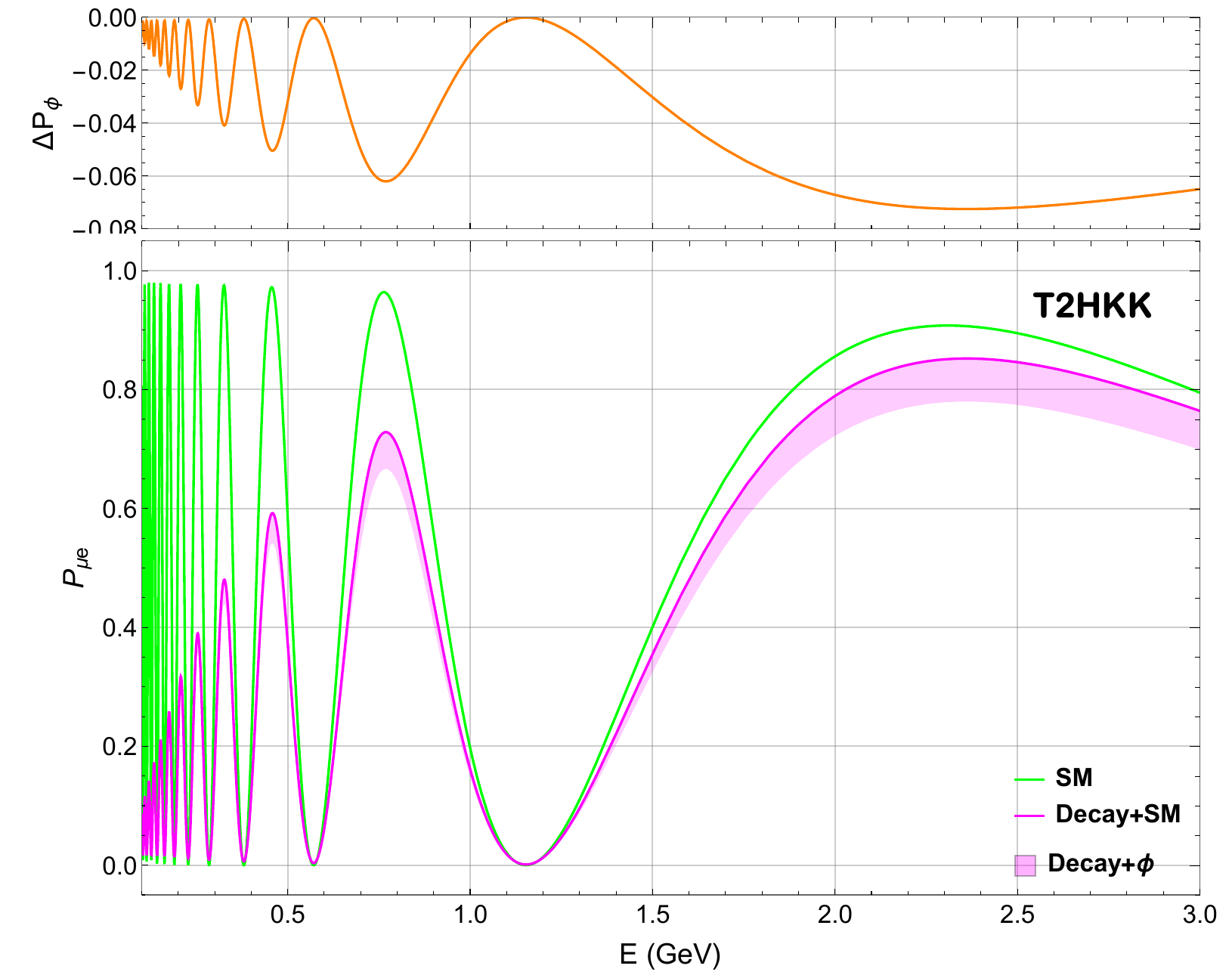}
\caption{The solid green line depicts the neutrino oscillation in matter, and the magenta line corresponds to the decay probability in the presence of SM matter effects. The light magenta band covers the maximum range of effects of non-zero values of Majorana phase $\phi$ in neutrino decay in matter. The orange solid line in the top subplot shows the maximum difference between the decay probability with and without the Majorana phase in the matter (denoted by $\Delta P_{\phi}$).} 
    \label{accelerator-w-majorana}
\end{figure*}

\begin{figure*}[htb!] 
    \centering
    \includegraphics[width = 0.47\textwidth]{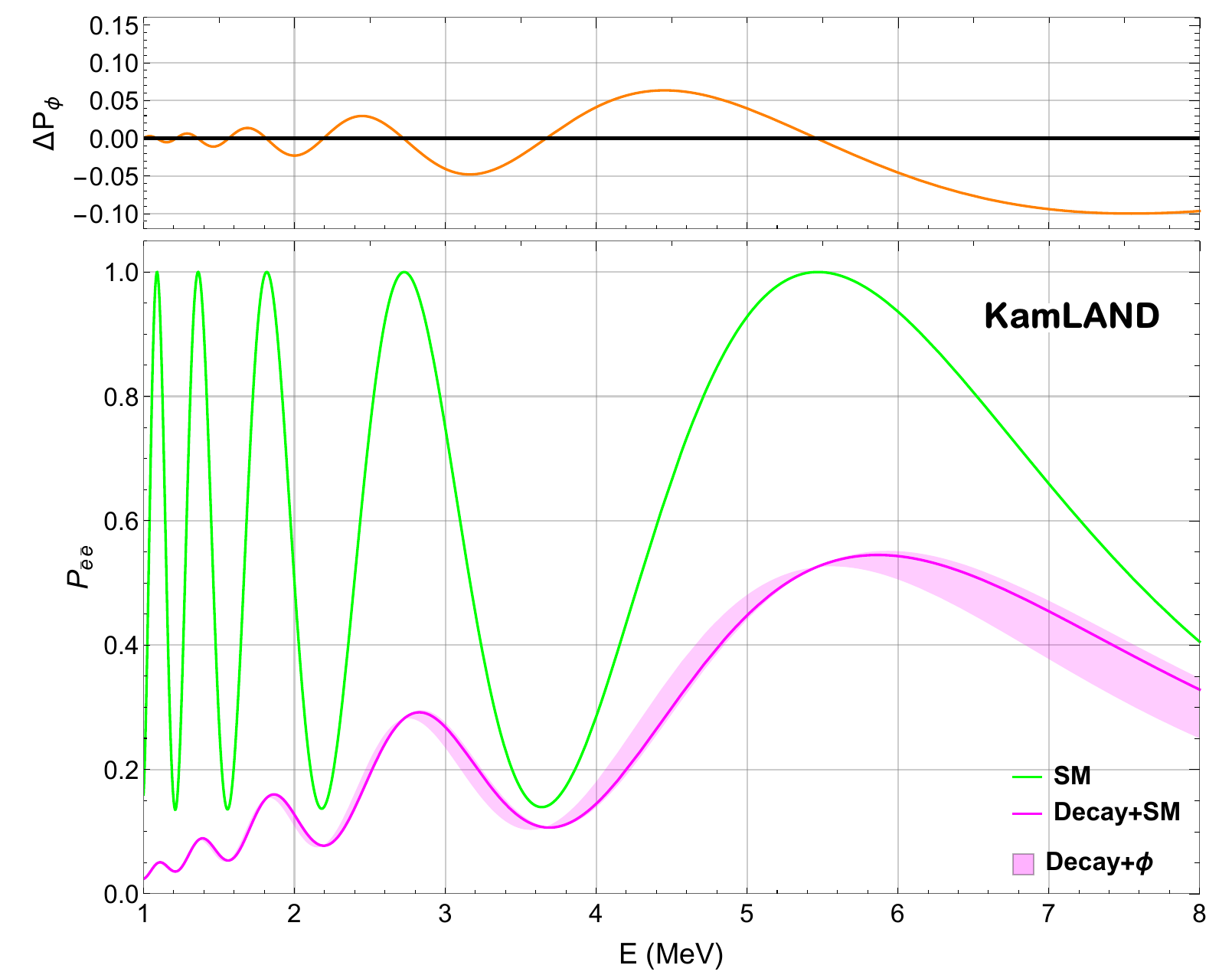}
    \includegraphics[width = 0.47\textwidth]{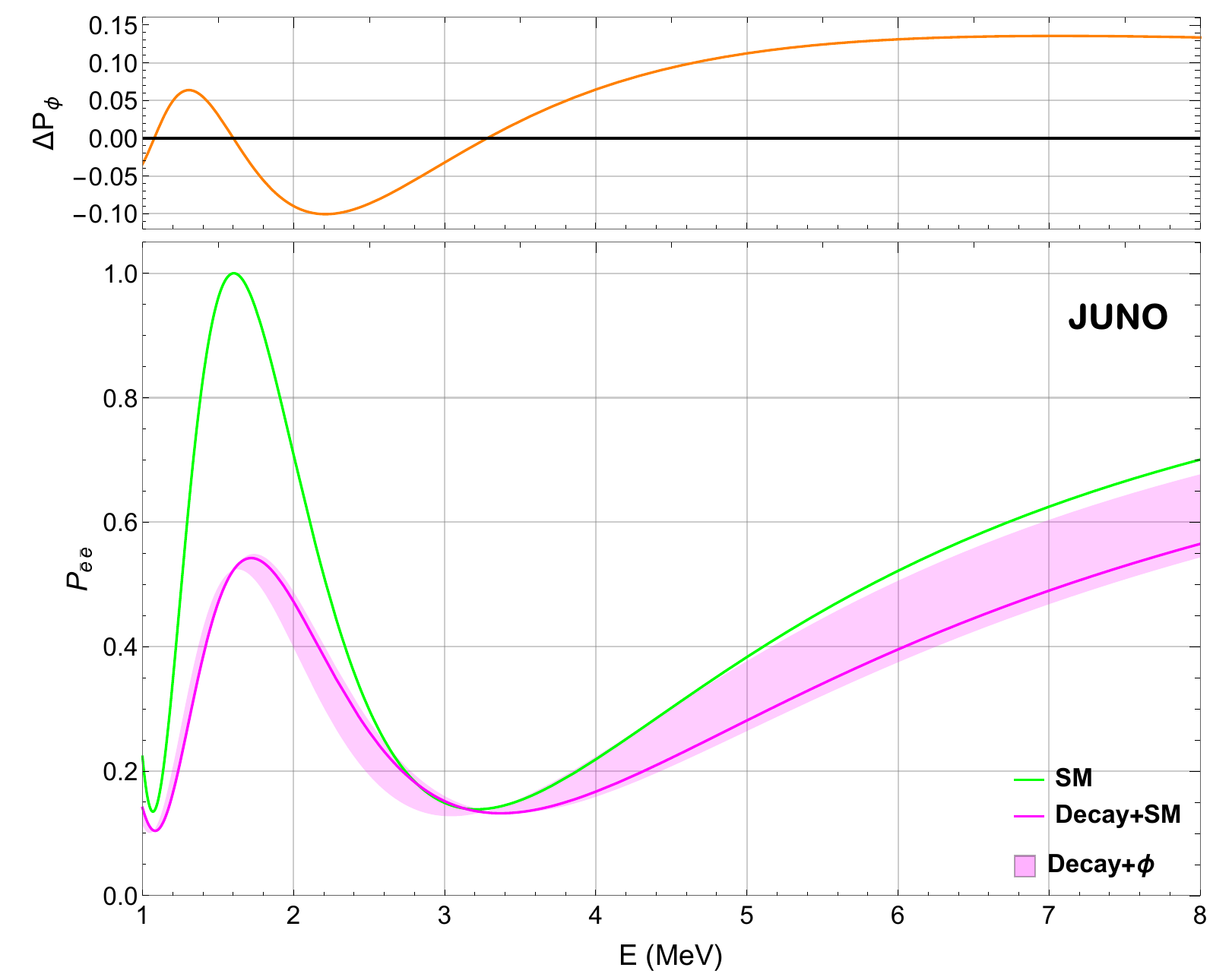}
    \caption{Survival probabilities for KamLAND and JUNO reactor experiments with Majorana phase. Here, the plots follow the same color code and definitions as in Fig. \ref{accelerator-w-majorana}}
    \label{reactor-w-majorana}
\end{figure*}

For accelerator neutrinos, we consider the experimental setups of DUNE \cite{DUNE:2015lol}, No$\nu$A \cite{NOvA:2004blv}, T2K \cite{T2K:2001wmr} and T2HKK\cite{Hyper-Kamiokande:2016srs}.  The DUNE experiment has a baseline of $L = 1300$ km and operates with a neutrino energy range of $E = (1 - 14)$ GeV, making it sensitive to the parameters $\theta_{23}$ and $\Delta m_{23}^{2}$. No$\nu$A, with a baseline of 810 km and an energy range of (1 - 5) GeV, and T2K/T2HK, with a 295 km baseline and an energy range of (0.1 - 6) GeV, are both sensitive to $\theta_{23}$ and $\Delta m_{23}^{2}$ parameters. The T2HK experiment has the same baseline as T2K. In this setup, however, a much larger water Cherenkov detector of volume 256 kton is utilised and also the beam power is increased as compared to T2K, providing better statistics  \cite{Hyper-Kamiokande:2018ofw}. The primary objective of this set-up is to study the CP asymmetry. With its baseline identical to T2K, the discussed results in this work for T2K also apply to T2HK.  
The T2HKK experiment is designed to function over a longer baseline of 1100 km and within an effective energy range of 0.1 to 3 GeV. Consequently, it is anticipated to observe substantial matter effects.

For reactor neutrinos, we consider the KamLAND \cite{KamLAND:2002uet} and JUNO \cite{JUNO:2015zny,JUNO:2021vlw} experiments. KamLAND operates with a baseline of $L = 180$ km and an anti-neutrino energy range of $E = (1 - 8)$ MeV. In contrast, JUNO has a baseline of 53 km and an anti-neutrino energy range between 1 to 8 MeV. Both KamLAND and JUNO experiments are sensitive to the parameters $\theta_{12}$ and $\Delta m_{21}^{2}$. Although the JUNO experiment is sensitive to the three-flavor oscillation parameters, we demonstrate the influence of the phases specifically in the context of two-flavor oscillations only.

Table \ref{NSI} encapsulates the values of mixing angles, mass squared differences, and the $1\sigma$ range of NSI parameters. In cases where the experimental setup is responsive to $\Delta m^{2}_{23}$, decay parameters are set as $(b_1,b_2,\gamma) = (3, 6, 8)\times 10^{-5}\, \rm{eV^2}/2E$. Conversely, if the experiment is sensitive to $\Delta m^{2}_{21}$, the parameters are chosen as $(b_1,b_2,\gamma) = (3, 6 ,8)\times10^{-6} \,\rm{eV^2}/2E$. Additionally, for both scenarios, $\chi$ is set to $\pi/4$ \cite{Chattopadhyay:2021eba}.

We first consider NSI effects on the neutrino transition probabilities for neutrino decay in matter. These effects for various accelerator and reactor neutrino experiments are shown in Figs. \ref{accelerator-wo-majorana} and \ref{reactor-wo-majorana}, respectively.  Observing the top-left panel of Fig. \ref{accelerator-wo-majorana}, it is evident that in the DUNE experimental configuration, the presence of neutrino decay in matter leads to a reduction in $(\nu_{\mu}-\nu_e)$ neutrino transition probability denoted as $P_{\mu e}^{\rm decay + SM}$, compared to the value $P_{\mu e}^{\rm SM}$ with only SM matter effects. This reduction is particularly pronounced in the energy range of (2-4) GeV. Conversely, introducing NSI effects in the absence of decay can either enhance or suppress the value of $P_{\mu e}$ (referred to as $P_{\mu e}^{\rm NSI}$). The curve for $P_{\mu e}$ in the decay-in-matter scenario falls within the band of $P_{\mu e}^{\rm NSI}$ which is due to the 1$\sigma$ allowed range of NSI parameters. This implies that a mere reduction in $P_{\mu e}$ cannot be unequivocally attributed to neutrino decay, as NSI effects can also induce such a depletion.

If we introduce NSI effects to $P_{\mu e}^{\rm decay + SM}$, it can lead to both suppression and enhancement in its value. The green lines in Fig.~\ref{accelerator-diffwo-majorana} represent the maximum deviation in $\Delta P_{\rm NSI}$ for DUNE, demonstrating that an enhancement of 5\% is possible for $E > 2$ GeV. Additionally, a suppression of more than 5\% is also possible across the entire energy range above 2 GeV, with a maximum depletion of 10\% occurring within the energy range of 2-3 GeV.

The figure also illustrates that the influence of pure NSI can be distinguished from all other considered effects if $P_{\mu e}$ is observed above the values predicted for $P_{\mu e}^{\rm SM}$, $P_{\mu e}^{\rm decay + SM}$, and $P_{\mu e}^{\rm decay + NSI}$. As mentioned earlier, any reduction in $P_{\mu e}$ may stem either from pure NSI effects or from decay in matter with SM. However, if $P_{\mu e}$ is depleted to a value as low as, for example, $\approx 0.7$ at $E \approx 2.5$ GeV, it can only be attributed to the combined effects of decay and NSI. Consequently, a precise measurement of $P_{\mu e}$ can facilitate the distinction between the pure NSI effect and the concurrent presence of decay and NSI in the DUNE experimental setup.

The conclusions remain nearly identical for the No$\nu$A setup. Here, it is observed that $P_{\mu e}^{\rm decay + NSI} < P_{\mu e}^{\rm decay + SM}$ for neutrino energies below 4 GeV, as indicated by the $\Delta {P}_{\rm NSI}$ plot. Moreover, a precise measurement of $P_{\mu e}$ in the energy range of 1-2 GeV holds the potential to yield signatures for the simultaneous presence of decay and NSI. For the T2K setup, $P_{\mu e}^{\rm decay + NSI} \approx P_{\mu e}^{\rm decay + SM}$. Like DUNE, the T2HKK experiment is also poised to enhance the discrimination of NSI and decay effects. Accurate measurements within the energy spectrum above 1.5 GeV can enable the differentiation between the concurrent effects of decay and NSI. The difference \(\Delta P_{\rm NSI}\), represented by a red line in Fig.~\ref{accelerator-diffwo-majorana}, indicates that an enhancement of up to 5\% is achievable in the 2-3 GeV energy range. Additionally, a suppression of approximately 5\%, with a peak depletion of 7\%, can occur within the same energy range for different sets of NSI parameters.

Let's now turn our attention to the reactor experiments KamLAND and JUNO. Examining the left panel of Fig. \ref{reactor-wo-majorana} for KamLAND, it becomes evident that the impact of NSI is noticeable only for neutrino energies above 6 MeV. In the case of neutrino decay in matter, $P_{\bar e\bar e}^{\rm decay + SM}$ is consistently lower than both $P_{\bar e\bar e}^{\rm SM}$ and $P_{\bar e\bar e}^{\rm NSI}$ across almost the entire considered energy range. Therefore, any observable reduction in $P_{\bar e\bar e}$ compared to the predicted values of $P_{\bar e\bar e}^{\rm SM}$ and $P_{\bar e\bar e}^{\rm NSI}$ can be attributed to neutrino decay. The introduction of NSI to neutrino decay can result in both enhancement and suppression in the value of $P_{\bar e\bar e}^{\rm decay + SM}$, as depicted in the plot of $\Delta P_{\rm NSI}$. However, the maximum enhancement can only reach up to 5\%, whereas the depletion is almost negligible.

Moreover, the influence of the combined presence of decay and NSI can be distinguished from other considered effects, as indicated in the figure, for neutrino energies around 8 MeV. Nevertheless, achieving this differentiation requires an extremely precise measurement of $P_{\bar e\bar e}$ as the effects are marginal. For the JUNO setup, the impact of NSI is minimal, with $P_{\bar e\bar e}^{\rm decay + SM}$ consistently lower than $P_{\bar e\bar e}^{\rm SM}$. Furthermore, the additional effect of NSI on decay proves to be too small to be observed.

We now explore the potential existence of the Majorana phase and analyze its impact on neutrino transition probabilities for neutrino decay in matter. The results for the considered accelerator and reactor experimental setups are illustrated in Figs. \ref{accelerator-w-majorana} and \ref{reactor-w-majorana}, respectively.
Examining the top left panel of Fig. \ref{accelerator-w-majorana}, it becomes apparent that, for the DUNE experiment, the presence of the Majorana phase can lead to the depletion of $P_{\mu e}^{\rm decay + SM}$ across the entire range of considered energy. This depletion is most pronounced in the energy range of 2-6 GeV. Interestingly, this effect  ($\Delta P_{\rm \phi} = P^{\rm decay + \phi} - P^{\rm decay + SM}$) mirrors the nature of depletion in $P_{\mu e}^{\rm decay + SM}$ caused by the presence of NSI ($\Delta P_{\rm NSI}$), as observed earlier. Thus, we observe that the effect of the Majorana phase on decay in matter with SM interaction can be replicated by the presence of NSI—indicating that the effect of NSI on decay can be comparable to the effect of the Majorana phase on decay. Furthermore, this conclusion holds true for the depletion in probabilities for the No$\nu$A, T2K and T2HKK experiments, as evident from the top right and bottom figures in ~\ref{accelerator-w-majorana}. For T2HKK, the depletion due to the presence of the Majorana phase is more evident for energies around $\sim 2$ \text{GeV}.

For reactor neutrino experiments, it is evident from Fig. \ref{reactor-w-majorana} that the Majorana phase does not appear in the neutrino oscillation probabilities in the presence of SM matter effects. For the KamLAND experiment, the presence of the Majorana phase suppresses the anti-neutrino survival probability $P^{\rm decay +\rm SM}_{\bar e \bar e}$ for the entire range of energies, which is opposite to the effect of NSI.
In the JUNO experimental setup, the presence of the Majorana phase can be realized only above 1.5 MeV, which is suppressing in nature up to 3 MeV and above that, it starts to enhance the $P^{ \rm decay + \rm SM}_{\bar e \bar e}$ for the entire considered energy range. Since the NSI effects are negligible in this setup, such enhancement and suppression can only be attributed to the additional phase.

The difference $\Delta P_\phi$  is depicted in the top-up subplots of individual reactor experiments in Fig. \ref{reactor-w-majorana}. For the KamLAND set-up, the value of $\Delta P_\phi$ can be as large as $\approx$ 10$\%$ for neutrino energies around 8 MeV and 16 MeV. However, a similar difference can appear for energies around 2 MeV and for energies above 5 MeV in the case of the JUNO experiment.

\begin{figure*}[htb]   
    \centering
    \includegraphics[width = 0.46\textwidth]{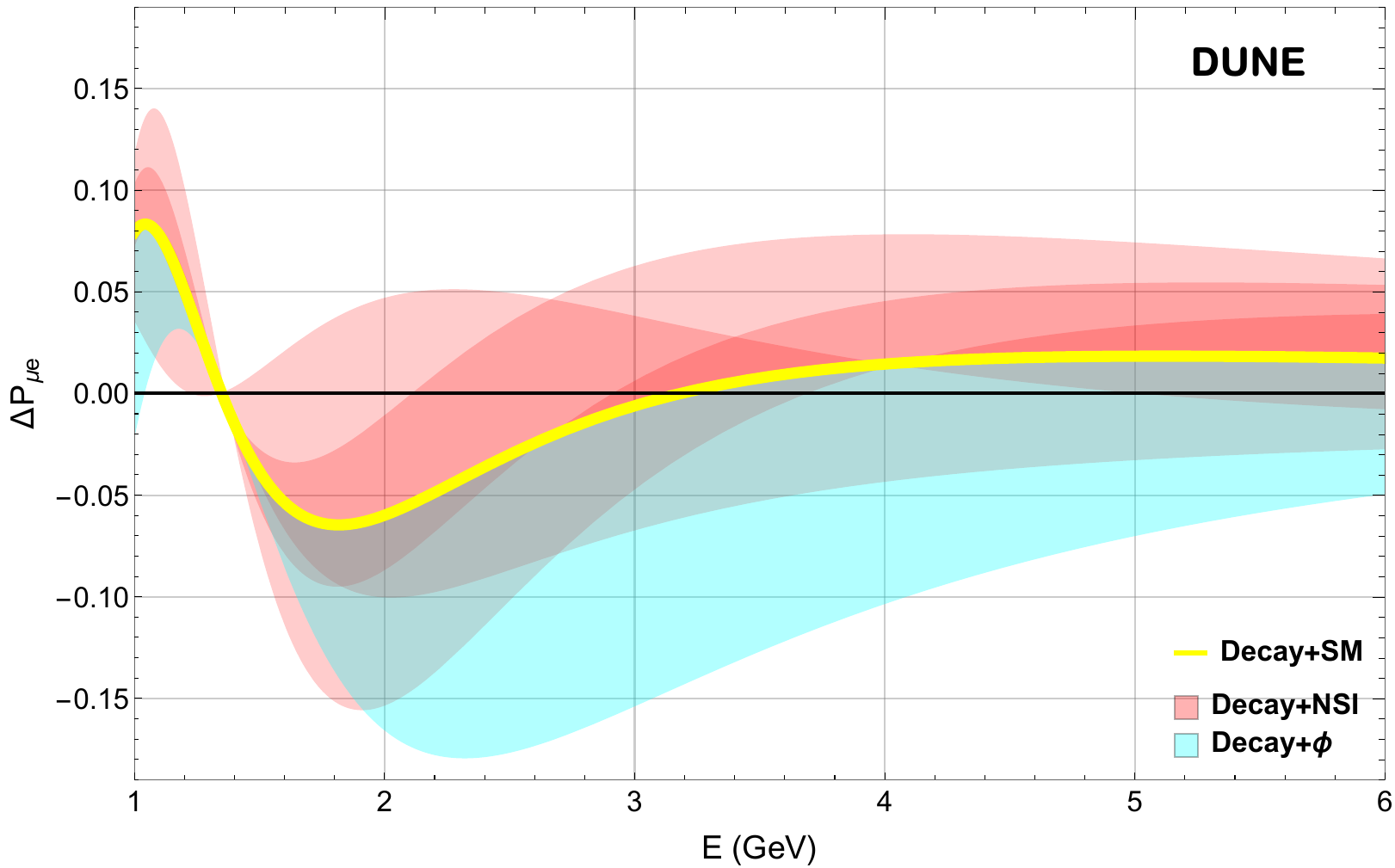}
    \includegraphics[width = 0.46\textwidth]{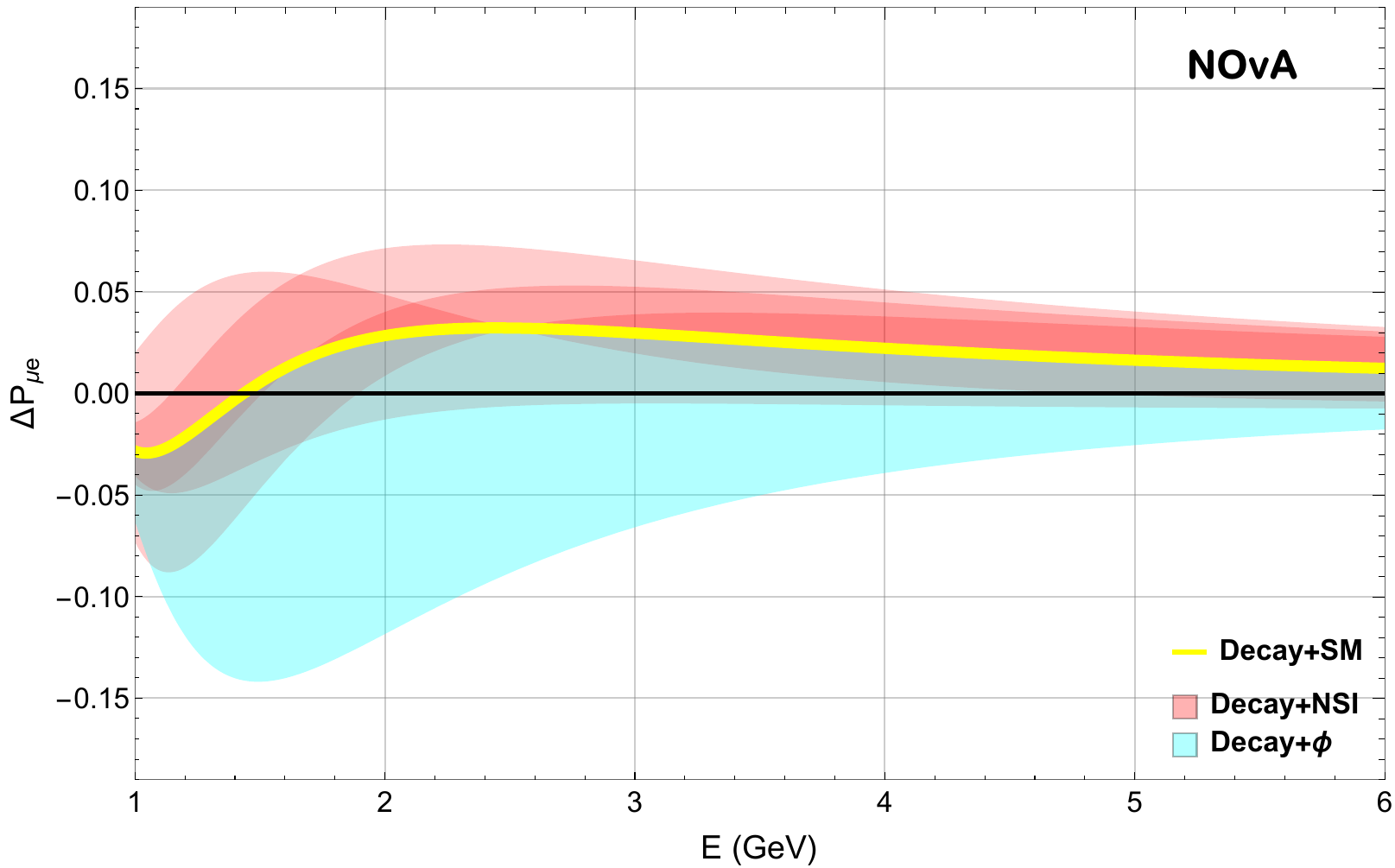}\\
    \includegraphics[width = 0.46\textwidth]{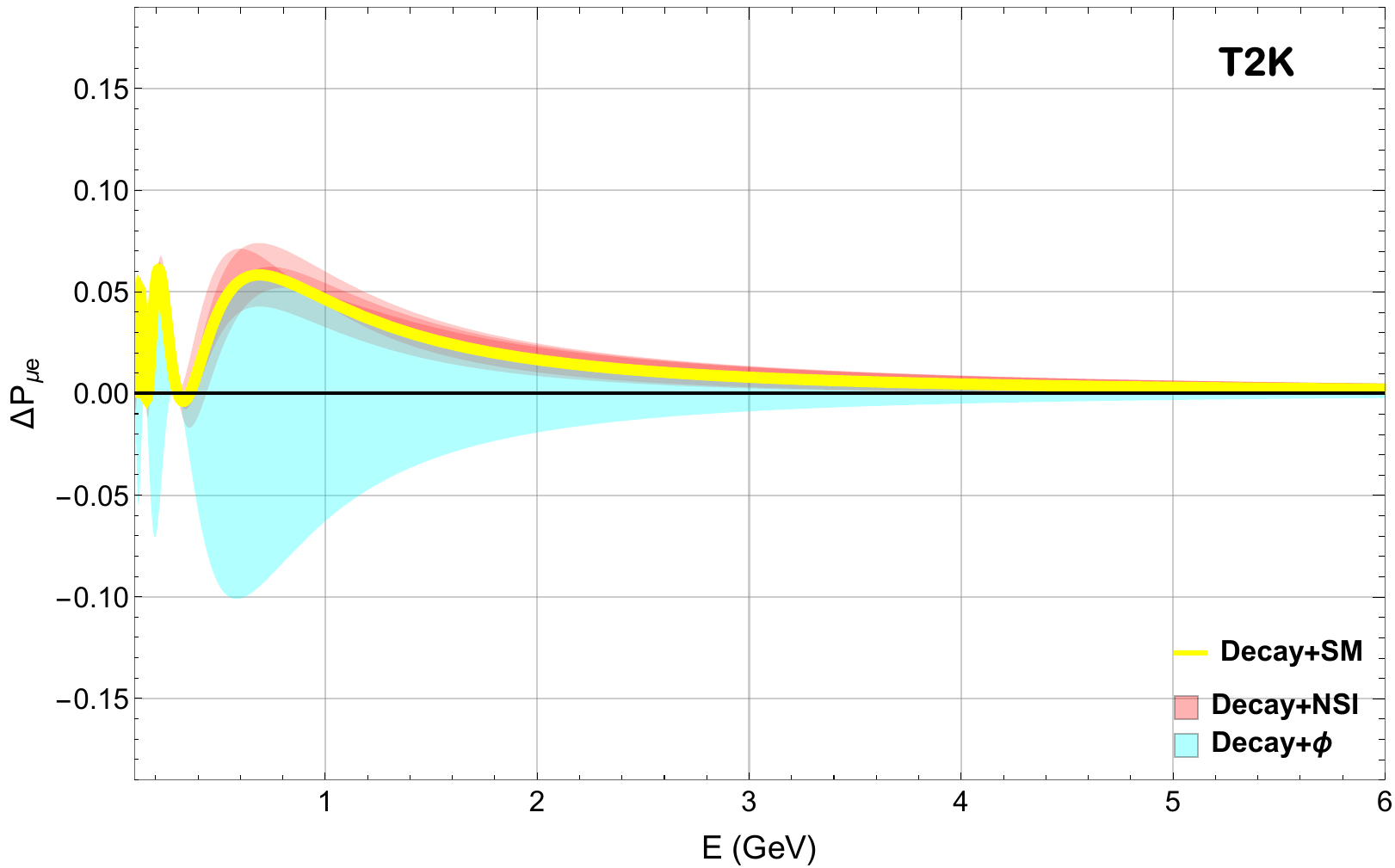} 
    \includegraphics[width = 0.46\textwidth]{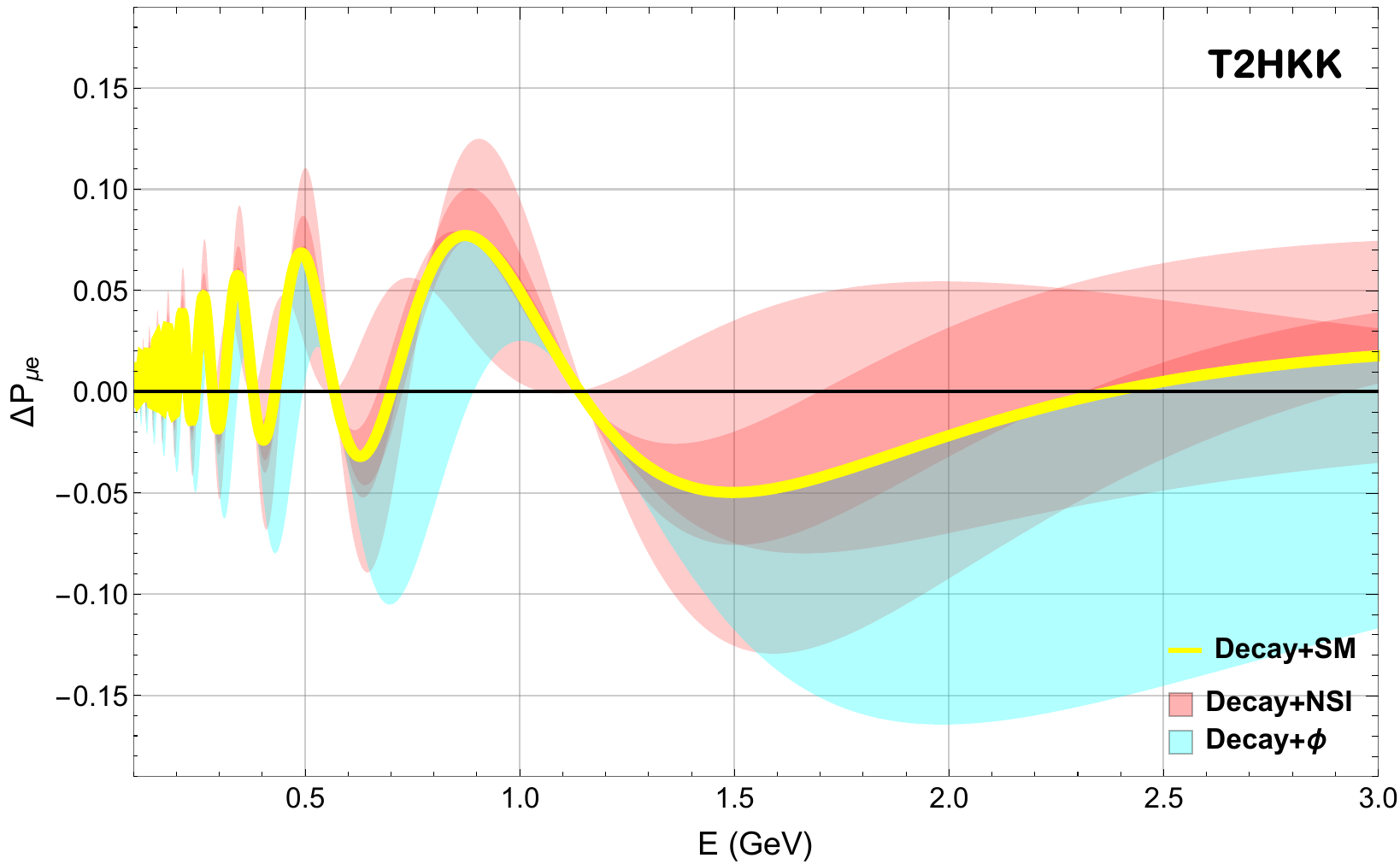} 
\caption{
The interplay of effects of Majorana phase and NSI parameters on $\Delta P_{\alpha\beta}$ are illustrated for the DUNE, No$\nu$A, T2K and T2HKK experimental setups. The yellow solid line depicts the difference $\Delta P_{\alpha\beta}$ in the decay probability within matter. The red-shaded region captures the impact of NSI on $\Delta P_{\alpha\beta}$, with varying shades representing different combinations of NSI parameters. Additionally, the blue-shaded region signifies the influence of a non-zero Majorana phase on $\Delta P_{\alpha\beta}$ for neutrinos undergoing decay in the presence of matter.}
    \label{deltaP:accelerator}
\end{figure*}

We now shift our focus to the observable $\Delta P_{\alpha\beta} = P_{\alpha\beta} - \Bar{P}_{\alpha\beta}$ with the aim of exploring additional avenues for observing and distinguishing between these subleading effects. We commence our analysis with the accelerator experiment DUNE, and the outcomes are illustrated in the top left panel of Fig. \ref{deltaP:accelerator}. Initially, we examine the results in the absence of the Majorana phase, i.e., for $\phi=0$. 
In the context of neutrino decay within matter with SM interactions, the maximum value of $\Delta P_{\alpha\beta}$ reaches approximately 0.8 at a neutrino energy of around 1 GeV. Beyond 1.5 GeV, the sign of $\Delta P_{\alpha\beta}$ turns negative. However, when both decay and NSI are present, the maximum value of $\Delta P_{\alpha\beta}$ further rises to approximately 0.15 at an energy of around 1 GeV. This specific scenario allows for such a high value below 1.5 GeV, setting it apart from other considered effects. Notably, the simultaneous presence of decay and NSI differentiates itself, as indicated by the positive values of $\Delta P_{\alpha\beta}$ between 1.5 and 3 GeV, a characteristic attributed solely to neutrino decay in the presence of NSI. In general, the values of $\Delta P_{\alpha\beta}$ for neutrino decay can consistently exceed the values observed for neutrino decay in matter, irrespective of the presence or absence of the Majorana phase, for specific sets of NSI parameters and this enhancement can reach up to 0.05 throughout the energy range above 2 GeV. Hence, a precise measurement of this observable holds the potential to distinguish the effects of neutrino decay with NSI from those of neutrino decay in matter.

To demonstrate the ability of $\Delta P_{\alpha\beta}$ to distinguish the effects of NSI and Majorana phase on $P^{\rm decay + SM}$, we consider $\phi = 0$ to $2\pi$. In general, the effect of the Majorana phase is to deplete the value of $\Delta P_{\alpha\beta}$ as compared to its value for neutrino decay in the matter. The influence of the phase $\phi$ becomes notably significant for $E > 1.5$ GeV, as evident from the plot. In this case, the magnitude of $\Delta P_{\alpha\beta}$ can be enhanced up to 0.18 for $E \approx 2.5$ GeV and remains above 0.10 in the energy range of 2.5 to 4 GeV. Hence, any measurement of $\Delta P_{\alpha\beta}$ exceeding 0.10 for neutrino energies greater than 3 GeV in the DUNE experimental setup can be attributed to the presence of the Majorana phase.

For the  No$\nu$A setup, the value of $\Delta P_{\alpha\beta}$ for neutrino decay in the presence of NSI can be larger than that of neutrino decay in matter for the entire energy range.  The detection of a positive $\Delta P_{\alpha\beta}$ value for $E< 1.5$ GeV can be attributed to the coexistence of decay and NSI. Additionally, any measurement of $\Delta P_{\alpha\beta}$ above 0.05 for neutrino energies beyond 1.5 GeV is indicative of this dual presence. In this setup, the effects of the Majorana phase can be readily distinguished from those of decay in the presence of NSI, as illustrated in the figure. 
The most striking feature is the possibility of the existence of large negative values of  $\Delta P_{\alpha\beta}$  in the energy range of (1 - 4) GeV. While the decay+NSI scenario also predicts negative values below 2.5 GeV, these values are comparatively smaller than the predictions for decay in the presence of the Majorana phase.

The T2K setup is also well-suited to capture the effects of the Majorana phase, as evidenced by the bottom row of Fig. \ref{deltaP:accelerator}. All considered scenarios with zero Majorana phase predict positive values of $\Delta P_{\alpha\beta}$ in almost the entire energy range. In contrast, all considered scenarios with non-zero Majorana phase allow for large negative values across the entire energy range. Thus, the sole measurement of $\Delta P_{\alpha\beta}$ with a negative value would provide evidence for a non-zero Majorana phase. 

In the T2HKK experiment, accurate measurements of \(\Delta P_{\alpha\beta}\) within the 2-3 GeV energy range are crucial for distinguishing NSI and Majorana effects amidst neutrino decay. Notably, the difference exceeds 15\% for the Majorana phase within this range, indicating its distinct presence. Additionally, at around 1 GeV, the combined effects of NSI and decay can boost \(\Delta P_{\alpha\beta}\) by up to 10\%, making their concurrent presence detectable. 
For the reactor neutrino experiments, electron antineutrino survival probabilities are measured. Therefore, $\Delta P_{\alpha\alpha}$ would not be able to capture the effects of complex phases like the Majorana phase $\phi$ and the decay phase $\chi$. The only difference in $\Delta P_{\alpha\alpha}$ is due to the matter effects. Therefore, in our analysis, we do not consider $\Delta P_{\alpha\alpha}$ for reactor neutrino experiments. 

 In this analysis, the impacts on oscillation probabilities are detailed for fixed energies. However, each experimental configuration has a defined limit to how precisely it can measure neutrino energies. These precision limitations may limit the ability within certain energy ranges to yield experimentally detectable effects, complicating the observation of predicted effects as well. It is crucial to verify whether the proposed scenarios can yield observable effects even when the energy distribution of neutrinos is considered.  To elaborate, imagine an experimental setup characterized by an energy resolution \(\delta E\) at a certain energy \(E\). Theoretically, this results in a range of probability values for standard scenarios (either without decay or NSI) spanning from \(E-\delta E\) to \(E+\delta E\), with the lowest and highest probabilities being \(P_{min}\) and \(P_{max}\), respectively. If a non-standard scenario produces a probability \(P'\) that falls within this range, it becomes challenging to discern whether \(P'\) is influenced by the non-standard scenario or merely reflects the finite energy resolution of the experiment. However, if \(P'\) is less than \(P_{min}\) or greater than \(P_{max}\) within the same energy range, such deviations could distinctly indicate the influence of a non-standard scenario. This principle also applies when comparing two different non-standard scenarios.\\
 For instance, consider the DUNE experiment (Fig. \ref{accelerator-wo-majorana}) having energy resolution $\sim 8\%$ at 3 GeV \cite{DeRomeri:2016qwo}.
 In this range, the oscillation probability, when considering neutrino decay alone, tends to cluster tightly around a value near \( \sim 0.8 \). When NSI is also considered, it introduces a wider range of possible probability values, though only a few of these overlap with the decay-only scenario. Importantly, there are probability values influenced by NSI that extend beyond what could be explained by the experimental setup's energy resolution alone. This indicates that such probabilities, which fall outside the expected range due to the finite resolution, are likely attributable directly to the effects of NSI rather than being artifacts of measurement limitations. 
 As indicated by Fig.\ref{accelerator-wo-majorana}, the oscillation probability value of \(0.7\) exceeds the range expected from the decay-only scenario, suggesting that this result can be ascribed only to the impact of NSI on neutrino decay processes.\\
The energy resolution in neutrino experiments, crucial for precise data analysis and interpretation, varies across different setups. The NOvA experiment exhibits an energy resolution between 9-14\% for electron neutrinos within the 0-5 GeV range, potentially influencing the detail with which neutrino oscillations are observed (\cite{NOvA:2021nfi}). For the T2HK experiment, the energy resolution can be approximately represented as $0.031\,E + 0.0822\,\sqrt{E}$ \cite{Barenboim:2024wdn}. The KamLAND experiment offers better resolution at lower energies, described as approximately \(6.4\% / \sqrt{E(\text{MeV})}\), enabling more precise lower-energy measurements (\cite{Decowski:2016axc}). Meanwhile, the JUNO experiment allows for an even superior resolution of approximately \(3\% / \sqrt{E(\text{MeV})}\), enhancing its capability to detect subtle neutrino interactions and properties (\cite{JUNO:2021vlw}). These varying levels of resolution are pivotal for the experiments' ability to accurately discern fine details in neutrino oscillation patterns and hence to  probve theoretical models effectively.\\
Also, the inability to measure the exact coherence among the mass eigenstates at production may modify the oscillation probability. This can affect the neutrino oscillation experiments as well. Such effects of finite wavepacket size and decoherence have been studied for both reactor and accelerator experimental setups such as JUNO \cite{Marzec:2022mcz}, KamLAND \cite{deGouvea:2021uvg}, T2K \cite{Gomes:2020muc}, DUNE and T2HK \cite{Barenboim:2024wdn}. The combination of the effect of decoherence with neutrino decay and NSI is expected to have some effects on the oscillation probabilities. However, the addition of neutrino decoherence to these effects is a non-trivial task and calls for a more rigorous analysis.
\section{Conclusion}
\label{concl}

The primary focus of this study is twofold. Firstly, to explore the ramifications of neutrino decay in the presence of NSI, i.e., when both decay and NSI effects coexist. Secondly, to investigate the influence of the Majorana phase on neutrino decay in matter. We examined these effects on the neutrino transition probabilities, denoted as $P_{\alpha \beta} \equiv  P(\nu_{\alpha} \to \nu_{\beta})$, as well as the difference $\Delta P_{\alpha \beta} \equiv P(\nu_{\alpha} \to \nu_{\beta}) - P(\bar {\nu}_{\alpha} \to \bar{\nu}_{\beta})$. These observables were scrutinized for both accelerator and reactor neutrino experimental setups. Our findings are summarized below:

\begin{itemize}
    \item The influence of the simultaneous presence of decay and NSI on $P_{\alpha \beta}$ can diverge from their individual effects within specific neutrino energy ranges in the DUNE, No$\nu$A and T2HKK experimental setups.

    \item The impact of the Majorana phase on decay in matter with SM interaction can be emulated by the presence of NSI. In other words, the effect of NSI on decay can be similar to the effect of the Majorana phase on decay.

    \item The measurement of substantial positive values of $\Delta P_{\alpha\beta}$  in the DUNE, No$\nu$A and T2HKK experiment may indicate the simultaneous presence of decay and NSI. Such an observation could serve as confirmation of the concurrent existence of these effects.

    \item Observing a significantly large negative value of $\Delta P_{\alpha\beta}$ in all accelerator experimental setups can be attributed to the presence of a non-zero Majorana phase. In particular, a single measurement of $\Delta P_{\alpha\beta}$ with a negative value in the T2K setup would serve as evidence for a non-zero Majorana phase.
    
\end{itemize}

Hence, precise measurements of the $P_{\alpha \beta}$ and $\Delta P_{\alpha\beta}$ observables have the potential to uniquely determine the presence of the Majorana phase by distinguishing its effects from the simultaneous presence of both neutrino decay and NSI.


\begin{thebibliography}{99}

\bibitem{DUNE:2015lol}
R.~Acciarri \textit{et al.} [DUNE],
`
[arXiv:1512.06148 [physics.ins-det]].

\bibitem{JUNO:2015zny}
F.~An \textit{et al.} [JUNO],
J. Phys. G \textbf{43}, no.3, 030401 (2016)
[arXiv:1507.05613 [physics.ins-det]].

\bibitem{JUNO:2021vlw}
A.~Abusleme \textit{et al.} [JUNO],
Prog. Part. Nucl. Phys. \textbf{123}, 103927 (2022)
[arXiv:2104.02565 [hep-ex]].


\bibitem{Ohlsson:2012kf}
T.~Ohlsson,
Rept. Prog. Phys. \textbf{76}, 044201 (2013)
[arXiv:1209.2710 [hep-ph]].

\bibitem{Miranda:2015dra}
O.~G.~Miranda and H.~Nunokawa,
New J. Phys. \textbf{17}, no.9, 095002 (2015)
[arXiv:1505.06254 [hep-ph]].

\bibitem{Proceedings:2019qno}
P.~S.~Bhupal Dev, K.~S.~Babu, P.~B.~Denton, P.~A.~N.~Machado, C.~A.~Arg\"uelles, J.~L.~Barrow, S.~S.~Chatterjee, M.~C.~Chen, A.~de Gouv\^ea and B.~Dutta, \textit{et al.}
SciPost Phys. Proc. \textbf{2}, 001 (2019)
[arXiv:1907.00991 [hep-ph]].

\bibitem{Farzan:2017xzy}
Y.~Farzan and M.~Tortola,
Front. in Phys. \textbf{6}, 10 (2018)
[arXiv:1710.09360 [hep-ph]].


\bibitem{Bakhti:2020fde}
P.~Bakhti and M.~Rajaee,
Phys. Rev. D \textbf{103}, no.7, 075003 (2021)
[arXiv:2010.12849 [hep-ph]].

\bibitem{Kopp:2007ne}
J.~Kopp, M.~Lindner, T.~Ota and J.~Sato,
Phys. Rev. D \textbf{77}, 013007 (2008)
[arXiv:0708.0152 [hep-ph]].


\bibitem{Bahcall:1972my}
J.~N.~Bahcall, N.~Cabibbo and A.~Yahil,
Phys. Rev. Lett. \textbf{28}, 316-318 (1972)

\bibitem{Acker:1991ej}
A.~Acker, S.~Pakvasa and J.~T.~Pantaleone,
Phys. Rev. D \textbf{45}, 1-4 (1992)

\bibitem{Acker:1993sz}
A.~Acker and S.~Pakvasa,
Phys. Lett. B \textbf{320}, 320-322 (1994)
[arXiv:hep-ph/9310207 [hep-ph]].

\bibitem{Berryman:2014qha}
J.~M.~Berryman, A.~de Gouvea and D.~Hernandez,
Phys. Rev. D \textbf{92}, no.7, 073003 (2015)
[arXiv:1411.0308 [hep-ph]].

\bibitem{Bandyopadhyay:2002qg}
A.~Bandyopadhyay, S.~Choubey and S.~Goswami,
Phys. Lett. B \textbf{555}, 33-42 (2003)
[arXiv:hep-ph/0204173 [hep-ph]].

\bibitem{Gonzalez-Garcia:2008mgl}
M.~C.~Gonzalez-Garcia and M.~Maltoni,
Phys. Lett. B \textbf{663}, 405-409 (2008)
[arXiv:0802.3699 [hep-ph]].

\bibitem{Frieman:1987as}
J.~A.~Frieman, H.~E.~Haber and K.~Freese,
Phys. Lett. B \textbf{200}, 115-121 (1988)

\bibitem{Gago:2017zzy}
A.~M.~Gago, R.~A.~Gomes, A.~L.~G.~Gomes, J.~Jones-Perez and O.~L.~G.~Peres,
JHEP \textbf{11}, 022 (2017)
[arXiv:1705.03074 [hep-ph]].

\bibitem{Ascencio-Sosa:2018lbk}
M.~V.~Ascencio-Sosa, A.~M.~Calatayud-Cadenillas, 
Eur. Phys. J. C \textbf{78}, no.10, 809 (2018)
[arXiv:1805.03279 [hep-ph]].

\bibitem{deGouvea:2015ndi}
A.~de Gouv\^ea and K.~J.~Kelly,
Nucl. Phys. B \textbf{908}, 318-335 (2016)
[arXiv:1511.05562 [hep-ph]].



\bibitem{Chattopadhyay:2021eba}
D.~S.~Chattopadhyay, K.~Chakraborty, A.~Dighe, S.~Goswami and S.~M.~Lakshmi,
Phys. Rev. Lett. \textbf{129}, no.1, 011802 (2022)
[arXiv:2111.13128 [hep-ph]].

\bibitem{Berryman:2014yoa}
J.~M.~Berryman, A.~de Gouv\^ea, D.~Hern\'andez and R.~L.~N.~Oliveira,
Phys. Lett. B \textbf{742}, 74-79 (2015)
[arXiv:1407.6631 [hep-ph]].

\bibitem{Chattopadhyay:2022ftv}
D.~S.~Chattopadhyay, K.~Chakraborty, A.~Dighe and S.~Goswami,
JHEP \textbf{01}, 051 (2023)
[arXiv:2204.05803 [hep-ph]].

\bibitem{Dixit:2022izn}
K.~Dixit, A.~K.~Pradhan and S.~U.~Sankar,
Phys. Rev. D \textbf{107}, no.1, 013002 (2023)
[arXiv:2207.09480 [hep-ph]].

\bibitem{Coloma:2020nhf}
P.~Coloma, I.~Esteban, M.~C.~Gonzalez-Garcia and J.~Menendez,
JHEP \textbf{08}, no.08, 030 (2020)
[arXiv:2006.08624 [hep-ph]].

\bibitem{NOvA:2004blv}
D.~S.~Ayres \textit{et al.} [NOvA],
[arXiv:hep-ex/0503053 [hep-ex]].

\bibitem{T2K:2001wmr}
Y.~Itow \textit{et al.} [T2K],
[arXiv:hep-ex/0106019 [hep-ex]].

\bibitem{Hyper-Kamiokande:2016srs}
K.~Abe \textit{et al.} [Hyper-Kamiokande],
PTEP \textbf{2018}, no.6, 063C01 (2018)
[arXiv:1611.06118 [hep-ex]].

\bibitem{Hyper-Kamiokande:2018ofw}
K.~Abe \textit{et al.} [Hyper-Kamiokande],
[arXiv:1805.04163 [physics.ins-det]].

\bibitem{KamLAND:2002uet}
K.~Eguchi \textit{et al.} [KamLAND],
Phys. Rev. Lett. \textbf{90}, 021802 (2003)
[arXiv:hep-ex/0212021 [hep-ex]].

\bibitem{DeRomeri:2016qwo}
V.~De Romeri, E.~Fernandez-Martinez and M.~Sorel,
JHEP \textbf{09}, 030 (2016)
[arXiv:1607.00293 [hep-ph]].

\bibitem{NOvA:2021nfi}
M.~A.~Acero \textit{et al.} [NOvA],
Phys. Rev. D \textbf{106}, no.3, 032004 (2022)
[arXiv:2108.08219 [hep-ex]].

\bibitem{Barenboim:2024wdn}
G.~Barenboim, A.~M.~Calatayud-Cadenillas, A.~M.~Gago and C.~A.~Ternes,
Phys. Lett. B \textbf{852}, 138626 (2024)
[arXiv:2402.16395 [hep-ph]].

\bibitem{Decowski:2016axc}
M.~P.~Decowski [KamLAND],
Nucl. Phys. B \textbf{908}, 52-61 (2016)


\bibitem{Marzec:2022mcz}
E.~Marzec and J.~Spitz,
Phys. Rev. D \textbf{106}, no.5, 053007 (2022)
[arXiv:2208.04277 [hep-ph]].

\bibitem{deGouvea:2021uvg}
A.~de Gouv\^ea, V.~De Romeri and C.~A.~Ternes,
JHEP \textbf{06}, 042 (2021)
[arXiv:2104.05806 [hep-ph]].

\bibitem{Gomes:2020muc}
A.~L.~G.~Gomes, R.~A.~Gomes and O.~L.~G.~Peres,
JHEP \textbf{10}, 035 (2023)
[arXiv:2001.09250 [hep-ph]].


\end{thebibliography}
\end{document}